\documentclass{ptephy}

\usepackage{boites}
\usepackage{natbib}
\setcitestyle{numbers}
\usepackage{subfigure}
\newlength{\subfigwidth}
\newlength{\subfigcolsep}
\usepackage{textcomp}

\begin{document}

\title{Study of hadron interactions in a lead-emulsion target}
\author{\name{\fname{Hirokazu}~\surname{Ishida}}{1}, \name{\fname{Tsutomu}~\surname{Fukuda}}{1},
 \name{\fname{Takafumi}~\surname{Kajiwara}}{1}, \name{\fname{Koichi}~\surname{Kodama}}{2},
 \name{\fname{Masahiro}~\surname{Komatsu}}{3}, \name{\fname{Tomokazu}~\surname{Matsuo}}{1},
 \name{\fname{Shoji}~\surname{Mikado}}{4}, 
 \name{\fname{Mitsuhiro}~\surname{Nakamura}}{3},
 \name{\fname{Satoru}~\surname{Ogawa}}{1},
 \name{\fname{Andrey}~\surname{Sheshukov}}{5}, \name{\fname{Hiroshi}~\surname{Shibuya}}{1},
 \name{\fname{Jun}~\surname{Sudou}}{1}, \name{\fname{Taira}~\surname{Suzuki}}{1},
 \name{\fname{Yusuke}~\surname{Tsuchida}}{1}}
\address{\affil{1}{Department of Physics, Toho University, Funabashi 274-8510, Japan}
\affil{2}{Aichi University of Education, Kariya 448-8542, Japan}
\affil{3}{Nagoya University, Nagoya 464-8602, Japan}
\affil{4}{Nihon University, Narashino 275-8576, Japan}
\affil{5}{JINR - Joint Institute for Nuclear Research, Dubna 141980, Russia}
\email{shibuya@ph.sci.toho-u.ac.jp}}

\begin{abstract}
Topological and kinematical characteristics of hadron interactions have been 
studied using a lead-emulsion target exposed to 2, 4 and 10 GeV/$c$ hadron beams.
A total length of 60 m $\pi^-$ tracks was followed using a high speed automated emulsion
scanning system. A total of 318 hadron interaction vertices and their secondary charged
particle tracks were reconstructed. Measurement results of interaction lengths, 
charged particle multiplicity,
emission angles and momenta of secondary charged particles are compared 
with a Monte Carlo simulation and appear to be consistent.
Nuclear fragments emitted from interaction vertices were also detected
by a newly developed emulsion scanning system with wide-angle acceptance.
Their emission angle distributions are in good agreement with the simulated distributions.
Probabilities of an event being associated with at least one fragment track
are found to be greater than 50$\%$ for beam momentum $P > 4$ GeV/$c$ and 
are well reproduced by the simulation. 
These experimental results validate estimation of the background due to hadron interactions 
in the sample of $\tau$ decay candidates in the OPERA $\nu_{\mu} \to \nu_{\tau}$
oscillation experiment.

\end{abstract}

\subjectindex{C30, C32, H14, H16}

\maketitle

\section{Introduction and overview}

\begin{figure}[htbp]
	\begin{center}
		\includegraphics[height=5cm]{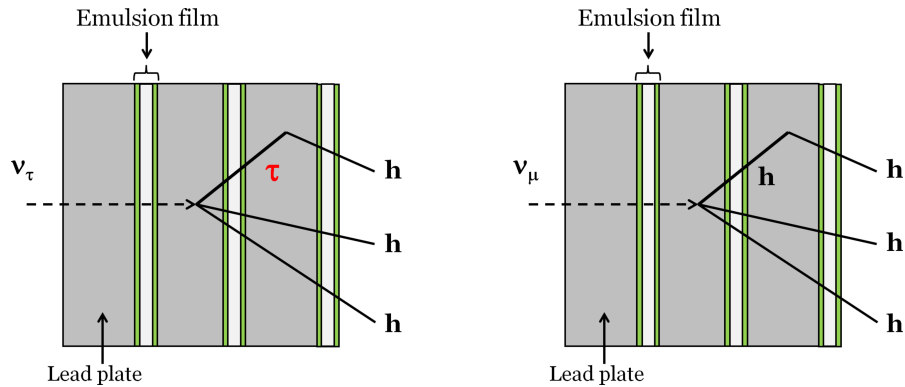}
		\caption[Comparison of $\tau $ decay with hadron interaction]
{Left: a hadronic decay of the $\tau$ lepton produced in a $\nu_{\tau}$ charged current
interaction. Right: an interaction of a secondary hadron ({\bf h}) produced in a $\nu_{\mu}$ 
neutral current interaction.}
		\label{fig:figure_1}
	\end{center}
\end{figure}

The neutrino oscillation channel $\nu _{\mu } \to \nu _{\tau }$ is considered
to be the dominant underlying process in the atmospheric neutrino sector.
To prove this, the detection, in an initially pure $\nu _{\mu }$ beam, 
of the $\tau$ lepton produced in a $\nu _{\tau }$ charged
current interaction is essential and provides a clear signature of the oscillation.
The nuclear emulsion is the only particle detector which can detect $\tau$ decays
by resolving their short flight path~\citep{DONUT_PLB,DONUT_NIM}.
The ongoing $\nu _{\mu } \to \nu _{\tau }$ oscillation experiment, 
OPERA~\citep{OPERA_proposal,OPERA_status,OPERA_detector}, employs the emulsion cloud chamber (ECC) technique
to provide both a large target mass and an excellent spatial resolution.
Furthermore, $\nu _{\tau }$ appearance can be used to probe new physics, 
such as non-standard interactions and light sterile neutrino 
scenarios~\citep{New_physics_Yasuda,New_physics_Ota}.
An ECC brick is composed of layers of emulsion films interleaved with heavy material plates.
In the ECC brick, a secondary hadron produced in neutrino interactions could interact with the
heavy materials and mimic a hadronic decay of the $\tau$ lepton,
as shown in Figure~\ref{fig:figure_1}.
Hence such a secondary hadron interaction could be a source of background for $\tau$ decays.
Expected number of hadron interactions as background in the sample of $\tau$ decay candidates
is evaluated with a Monte Carlo simulation,
which must be validated by experimental data~\footnote{
The momentum region relevant to the hadron interaction background in the OPERA experiment
is around 2~GeV/$c$ to 10~GeV/$c$.}.

A high speed automated microscope system, S-UTS~\citep{SUTS},
has been developed to analyze particle tracks in emulsion films.
This system allows to follow beam particle tracks for a long distance and measure 
their interactions with large statistics. In addition, a new type of automated emulsion
scanning system~\citep{fukuda_sys11} has recently been developed 
to detect nuclear fragments emitted with large angles.

We have studied hadron interactions in an ECC brick exposed to 2, 4 and 10 GeV/$c$ hadron beams
and have compared experimental results with those of the Monte Carlo simulation.

\section{Detectors and beam exposure}

\begin{figure}[hbp]
	\begin{center}
		\begin{tabular}{c}
		
		\begin{minipage}{0.47\hsize}
			\begin{center}
				\includegraphics[height=4.5cm]{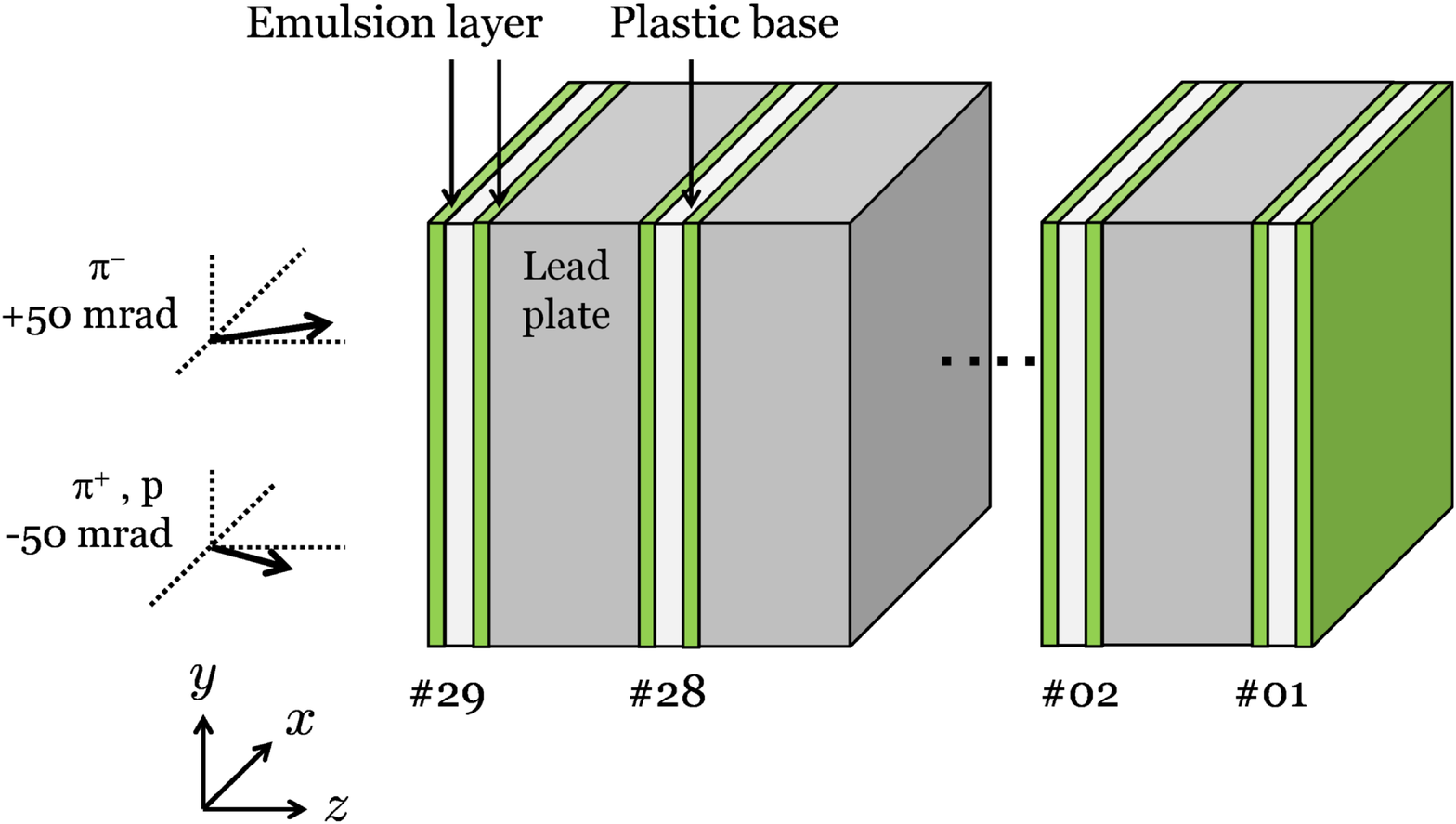}
			\end{center}
		\end{minipage}
	
		\begin{minipage}{0.47\hsize}
			\begin{center}
				\includegraphics[height=5cm]{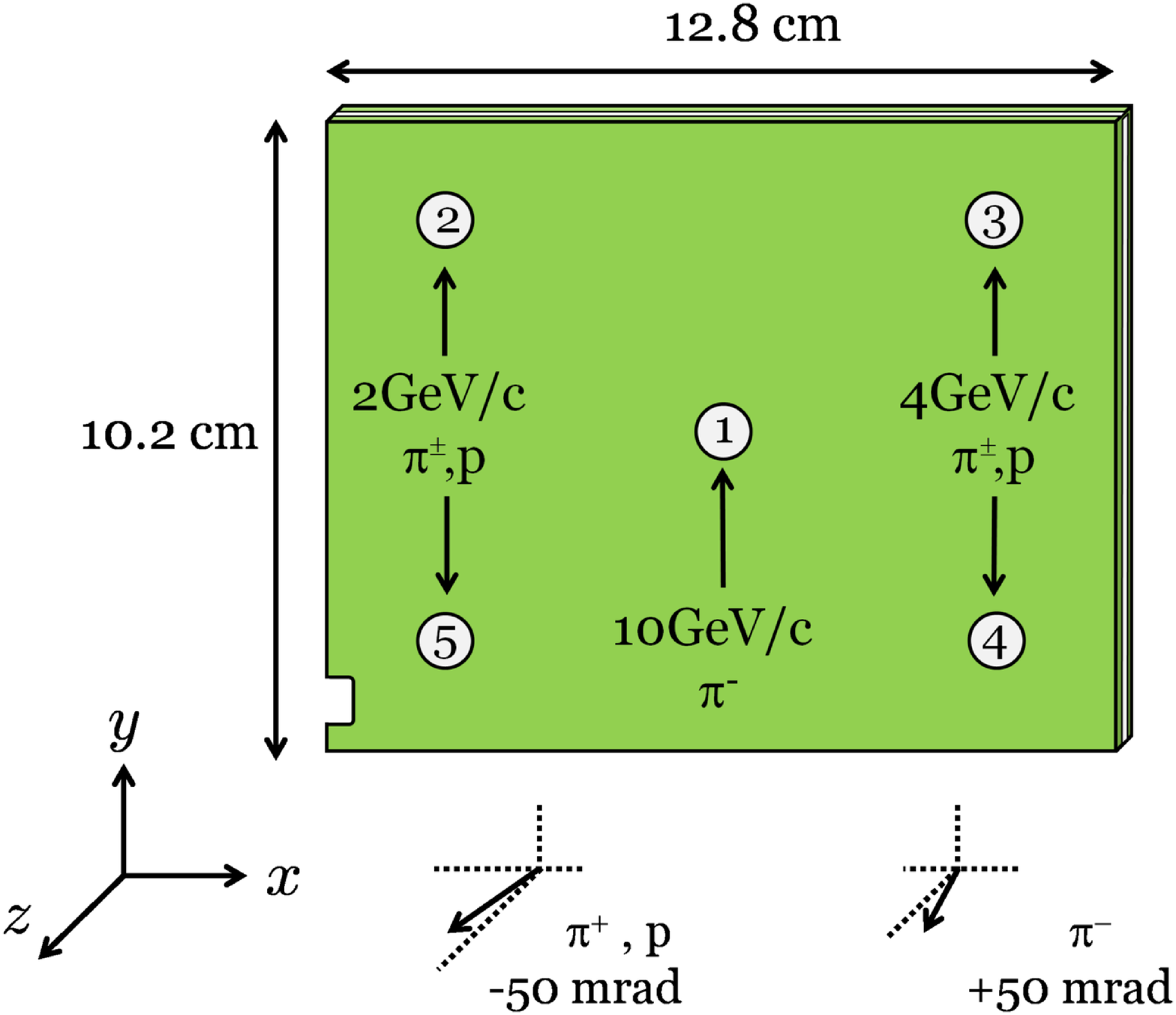}
			\end{center}
		\end{minipage}
	\end{tabular}
	\end{center}
		\caption[CERN ECC setup]{Left: schematic view of the ECC brick structure. 
Right: beam spot positions on the ECC brick.}
		\label{fig:figure_2}
\end{figure}

We exposed an ECC brick to 2, 4, 10~GeV/$c$ secondary hadron beams
at CERN PS-T7 beam line in May 2001.
The brick was composed of 29 emulsion films~\citep{OPERAfilm} 
(44~\textmu m \;thick emulsion layers on both sides of a 205~\textmu m \;thick plastic base) 
interleaved with 28 lead plates, 1~mm \;thick.
The brick was 12.8 cm wide, 10.2 cm high and 3.7 cm thick.
It was put on a turntable and tilted in the horizontal plane 
by an angle with respect to the beam of $\pm 50$~mrad as shown in Figure~\ref{fig:figure_2}.
The relative beam momentum spread, $\Delta P/P$, is $1 \%$~\citep{PS}.
The negatively (positively) charged beams are mainly composed of pions (pions and protons) 
with small contamination from electrons (positrons).
The exposure conditions; composition, momentum, tilt angle, spot size,
mean density integrated over the spot size are summarized in 
Table~\ref{table:cern_exposure}. 

\begin{table}[t]
		\caption[CERN ECC beam exposure]{Summary of beam exposures. 
For each beam spot, its composition, its momentum $P$, its angle with respect to the perpendicular
of the emulsion film, its spot size, and its mean density integrated over the spot size are tabulated.
For each spot, a tilt of $-50$ mrad corresponds to the positively charged hadron beam ($\pi^+, p$) 
and $+50$ mrad to the negatively charged hadron beam ($\pi^-$).
}
		\label{table:cern_exposure}
		\begin{center}
		\begin{tabular}{cccccc} \hline \hline
			Spot & Composition & $P$ [GeV/$c$] & Angle [mrad] & Spot size $\phi $ [mm] & Mean density [/cm$^2$] \\ \hline \hline
			1 & $\pi ^-$ & 10 & $+$ 50 & $\sim 3.0$ & $3.0 \times 10^3$ \\ \hline
			2 & $\pi ^\pm, p $ & 2 & $\mp $ 50 & $\sim 10.0$ & $14.0 \times 10^3$ \\ \hline
			3 & $\pi ^\pm, p $ & 4 & $\mp $ 50 & $\sim 6.0$ & $10.0 \times 10^3$ \\ \hline
			4 & $\pi ^\pm, p $ & 4 & $\mp $ 50 & $\sim 7.5$ & $1.5 \times 10^3$ \\ \hline
			5 & $\pi ^\pm, p $ & 2 & $\mp $ 50 & $\sim 15.0$ & $2.0 \times 10^3$ \\ \hline \hline
		\end{tabular}
	\end{center}
\end{table}

\section{Measurement and analysis method}

\subsection{Beam tracks}

We scanned the whole area of all the emulsion films by using S-UTS. 
Firstly track segments, so called micro tracks, were detected on each layer of emulsion films.
Positions ($x, y$) and slopes ($\tan \theta_x, \tan \theta_y$) of the micro tracks were measured.
Base tracks were reconstructed by connecting two corresponding micro tracks 
(reconstructed in emulsion layers) across the plastic base.
The slope acceptance of S-UTS is $\tan \theta < 0.6$, where $\theta$ is the track angle with 
respect to the perpendicular of the emulsion film (the $z$ axis). The track finding efficiency,
the probability to find a base track in a film, is evaluated to be $94.8 \pm 0.2 \%$
by examining whether a base track exists or not in the middle of 5 consecutive films.

The positions ($x, y$) and slopes ($\tan \theta_x, \tan \theta_y$) 
of base tracks were measured with respect to the microscope's coordinate system.
After connection of base tracks among emulsion films are performed, 
we adjust rotation, slant, parallel translation and distance 
between every pair of adjacent two emulsion films, 
so that differences of track positions and slopes in two films are minimized.   
Residuals of positions and slopes are $\sim 4$~\textmu m and $\sim 0.004$ respectively.

Good quality beam tracks are selected by requiring they have track segments found
in all 3 most upstream films and their slope 
in the most upstream film is compatible
with the average slope within 0.004 (1 $\sigma$)
where the average is calculated from all the measured beam track slopes.
For further analysis to be compared with the simulation,
good negatively charged pion ($\pi^-$) tracks with these conditions 
were selected in the area of $1.0 \times 1.0$~cm$^2$
from 1st, 4th and 5th spots~\footnote{
The 2nd and 3rd beam spots were not used because their mean densities were 
too high for the analysis.}, each corresponding to 10, 4 and 2~GeV/$c$,  
as shown in Figure~\ref{fig:figure_2}.
On the other hand, positively charged hadron beams are mainly composed of $\pi^+$s and protons,
which cannot be differentiated in this momentum range. Relative abundance is not known precisely. 
Therefore positively charged hadrons are not easy to be compared with the simulation.

\subsection{Interactions}
The beam tracks in the ECC brick 
are composed of successive track segments and some of them are missing 
due to inefficiency or hadron interactions.
The reconstructed tracks are followed down until no track segments
found in 3 consecutive films (Figure~\ref{fig:figure_3}, Left),
then a beam particle is supposed to interact at the lead plate just downstream the film
where the last track segment is found. 
The last track segment and its extrapolated position in the film 
downstream the supposed interaction position, the vertex film, are visually inspected.
If the last track segment is confirmed that it is a real track
and no track is found in the vertex film,
the interaction is then considered as confirmed.
If a track exists near the predicted position in the vertex film, we measure its position
and slopes manually and the beam track is followed further downstream. 
For beam tracks escaping from the downstream face of the brick, 
a straight line fit is applied to detect a kink topology somewhere in the brick. 
If a kink with $\theta _{\mbox{kink}} > 20$~mrad
is found, it is also considered as an interaction.

\subsection{Secondary particles}

\begin{figure}[htbp]
	\begin{center}
		\begin{tabular}{c}
		
		\begin{minipage}{0.45\hsize}
			\begin{center}
				\includegraphics[height=5cm]{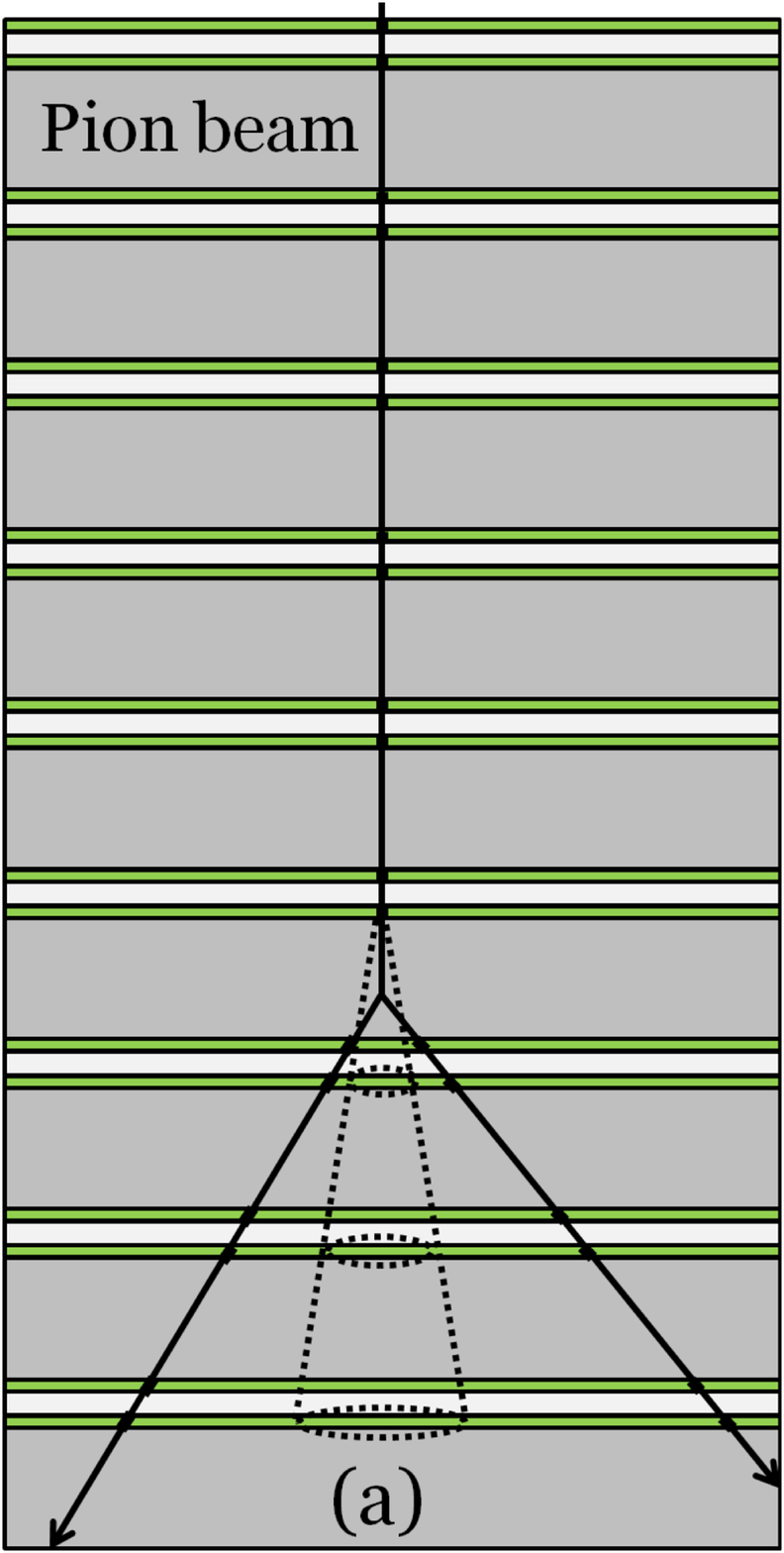}
			\end{center}
		\end{minipage}
	
		\begin{minipage}{0.45\hsize}
			\begin{center}
				\includegraphics[height=5cm]{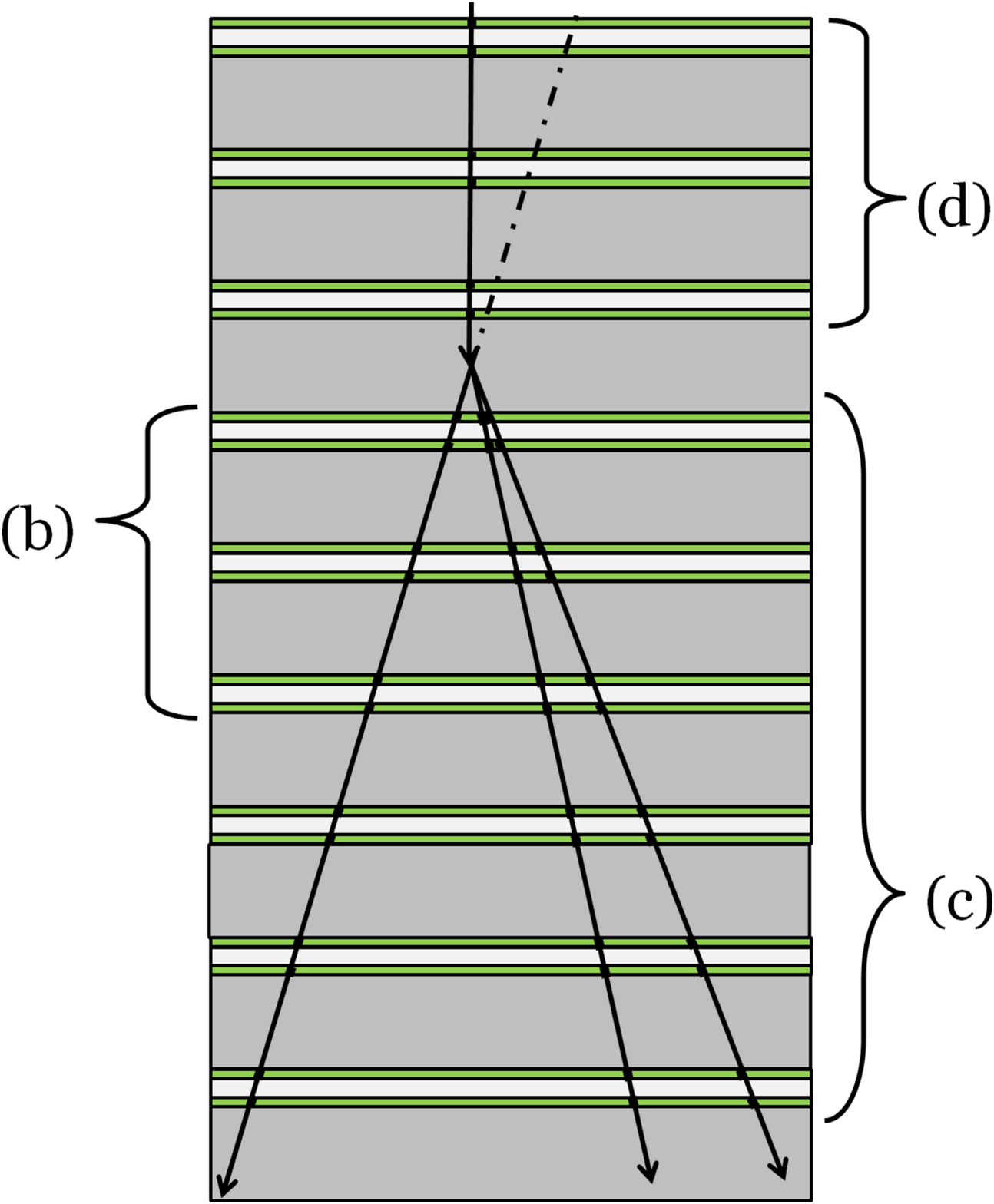}
			\end{center}
		\end{minipage}
	\end{tabular}
	\end{center}
	\caption[Detection method of beam interactions and of secondary particles]
{Left: detection method of beam interactions. 
Beam tracks are followed down until they are not found in three consecutive films. 
Extrapolated area (a), where a matching track is sought,
becomes larger in the downstream films.
Right: detection method of secondary particles. 
(b) A base track which satisfies the condition, $MD < 
(10+ \; 0.01 \times \Delta z)$~\textmu m,
must exist in at least 1 of the first 3 downstream films. 
(c) The base track must exist in at least 3 out of the 6 downstream films. 
(d) If the track exists in $\geq$ 2 out of 3 upstream films, 
it is considered as a passing-through track.}
	\label{fig:figure_3}
\end{figure}

Each confirmed interaction is subject to the secondary particle track search
inside the area of $3.0 \times 3.0$~cm$^2$ in 6 downstream films of the interaction vertex. 
Secondary particle tracks are sought under the following conditions,
as shown in Figure~\ref{fig:figure_3} right.

The angular acceptance of secondary particle track search is $\tan \theta < 0.6$, 
which is limited by the ordinary S-UTS performance.
For a secondary particle candidate found in at least 1 of the first 3 downstream films
including the vertex film, 
we require the minimum distance $MD$ of the candidate track with respect to the parent beam track
to be less than ($10+ \; 0.01 \times \Delta z$)~\textmu m taking into account 
both the measurement error and the error due to the multiple Coulomb scattering~(MCS) of the tracks, 
where the $\Delta z$~(\textmu m) is the $z$ distance of the $MD$ position from the nearest emulsion layer.
The scanning data have a lot of fake tracks due to low energy Compton electrons and random noises.
In order to reject such fake tracks, we demand the condition that base tracks exist in $\geq 3$
out of the 6 downstream films and they can be reconstructed as a single secondary track.
If base tracks matching with the secondary track candidate 
exist in $\geq$ 2 films out of the 3 upstream films, 
it is considered as a passing-through track which is not relevant to the interaction vertex.

Once a secondary track candidate is found, it is checked by visual inspection in 3 films:
1 upstream and 2 downstream films of the interaction vertex, in order to eliminate the
possibility of a passing-through track or an electron track originated from $\gamma$ conversion
in a downstream lead plate. 
The efficiency of the secondary particle track search was evaluated to be 99.2\%
by applying the S-UTS track finding efficiency to the conditions
in the whole procedure described above. 

\begin{figure}[htbp]
	\begin{center}
		\includegraphics[height = 6cm]{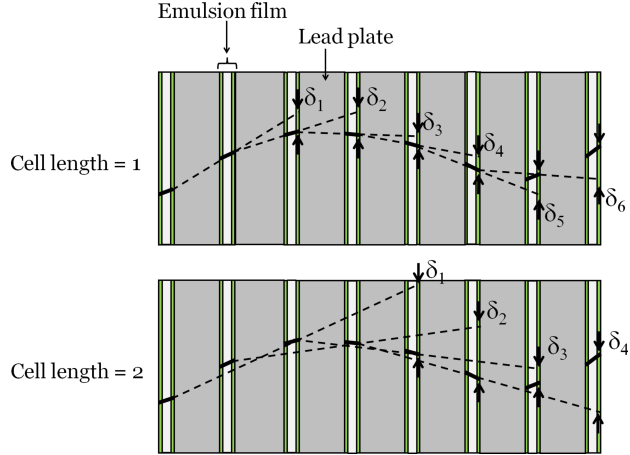}
		\caption[second difference]{Schematic view of the multiple Coulomb scattering (MCS) measurement.
The second differences of cell length = 1 and 2 are 
defined in top and bottom figures,
respectively.}
		\label{fig:figure_4}
	\end{center}
\end{figure}

Momenta of secondary tracks are estimated by measuring their multiple Coulomb scattering in the brick.
There are two methods in the multiple scattering measurements, the coordinate method~\citep{MCS_DONuT} 
and the angular method~\citep{MCS_OPERA}.
We use the coordinate method because it is more accurate when small multiple scattering signal is expected
although it requires precise alignment among the emulsion films.
We define the second difference $\delta_i$ in one projection by:
\begin{eqnarray}
   \delta_i =  x_{i+2} - x_{i+1} - \frac{x_{i+1} - x_{i}}{z_{i+1} - z_{i}}\cdot(z_{i+2} - z_{i+1})
\label{exp:2nd diff}
\end{eqnarray}
where the base track position at $i$th film ($i \in {1, . . . , n}$) is ($x_i, z_i$),
as shown in Figure~\ref{fig:figure_4}.
A similar expression can be written in the other projection, ($y_i, z_i$), and
data in the two projections are used as independent measurements of the second difference. 
The multiple scattering signal $\Delta_{\rm sig}$ is related to the momentum $p$ 
by~\citep{pdg,photomask}:
\begin{eqnarray}
	\Delta_{\rm sig} = \frac{t}{2\sqrt{3}} \frac{0.0136 \;{\rm GeV}}{p \, \beta \, c } 
\cdot \sqrt{\frac{t}{X_0}}\Bigl\{ 1 + 0.038 \ln \Bigl(\frac{t}{X_0}\Bigr) \Bigr\} 
\label{exp:delta x}
\end{eqnarray}
where $\beta c$, $t$, and $X_0$ are the particle velocity, distance traversed, and 
radiation length in the material respectively.
On the other hand, it is measured from second differences $\delta_i$ and their errors $\epsilon_i$:
\begin{eqnarray}
	<\delta_i^2> = \Delta_{\rm sig}^2 + <\epsilon_i^2>
\label{exp:sigma_x}
\end{eqnarray}
where $< >$ means the average over the available data.
Since the multiple scattering measurement is statistical estimation of the particle momentum,
it needs a sufficient number of data. To obtain reliable estimation,
we apply the coordinate method
only when at least 14 emulsion films are available for the position measurements.

\subsection{Nuclear fragments}
\begin{figure}[htbp]
	\begin{center}
		\includegraphics[height=5cm]{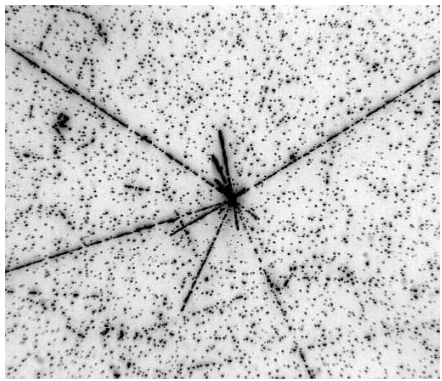}
		\caption[Hadron interaction]{Photograph of a hadron interaction in an emulsion layer. 
Nuclear fragments are observed as black or gray tracks.
Association of such highly ionizing particles is evidence of a hadron interaction. 
}
		\label{fig:figure_5}
	\end{center}
\end{figure}

If a secondary particle has a value of $\beta < 0.7$, 
the particle is observed as a heavily ionizing track or a nuclear fragment,
as shown in Figure~\ref{fig:figure_5}.
Nuclear fragments emitted from hadron interactions have also been sought
by a newly developed automatic emulsion scanning system~\citep{fukuda_sys11}.
The acceptance to the track slopes are much improved up to $|\tan \theta| < 3.0$.
Since nuclear fragments are emitted almost isotropically,
the new scanning system with a wider angular acceptance is suitable
in order to detect the nuclear fragments with a good efficiency.
The system has a large field of view, $352 \times 282$~\textmu m$^2$,
and has a track finding efficiency of practically 100\% for nuclear fragments~\citep{fukuda_sys11}.

Searched area is defined as $3.5 \times 2.5$~mm$^2$ of both upstream and downstream films of
each interaction vertex and its angular acceptance is $|\tan \theta| < 3.0$.
Since the nuclear fragments have low energy (few tens of MeV), they are expected to 
suffer large multiple scattering. Therefore the loose condition, 
$IP < (100+ \; 0.01 \times \Delta z)$~\textmu m 
is imposed.
Here the ``Impact Parameter" $IP$ is defined as the perpendicular distance between the path of a track
and the vertex which is reconstructed by the beam and secondary tracks.
The $\Delta z$~(\textmu m) is the distance between the vertex and the nearest emulsion layer
where the nuclear fragment track is measured.
A narrower condition, 
$IP< (50+ \; 0.01 \times \Delta z)$~\textmu m,
is imposed
in the case of 10~GeV/$c$ interactions 
because the beam density is high.
To remove fake tracks, each found track candidate is required to be a base track
and is confirmed by visual inspection.

\section{FLUKA simulation}

A FLUKA~\footnote{The FLUKA version 2011.1 was used in this study.}~\citep{FLUKA1,FLUKA2} based Monte Carlo simulation
was employed for comparison.
30,000 negatively charged beam pions each were generated for momentum of $P$ = 2, 4 and 10 GeV/$c$.
The generated beam pions, of which directions were spread using observed distributions in this experiment,
entered the brick. Some of them interacted with the brick material according to their cross sections
and produced secondary particles including nuclear fragments by using the PEANUT model~\citep{FLUKA_PEANUT}
of hadron interactions.
Exposed density, position and slope measurement errors, detection efficiency of base tracks
were adjusted to be the same as real experimental conditions.

Then the simulation analysis has been performed by the same methods and tools
that are used for the real data.
Secondary particles with $\beta \geq 0.7$ ($\beta < 0.7$) 
are defined as relativistic particles (non-relativistic particles, i.e. nuclear fragments).

\section{Results and discussion}

\subsection{Interaction length}

\begin{figure}[hbp]
	\begin{center}
		\includegraphics[height=7cm]{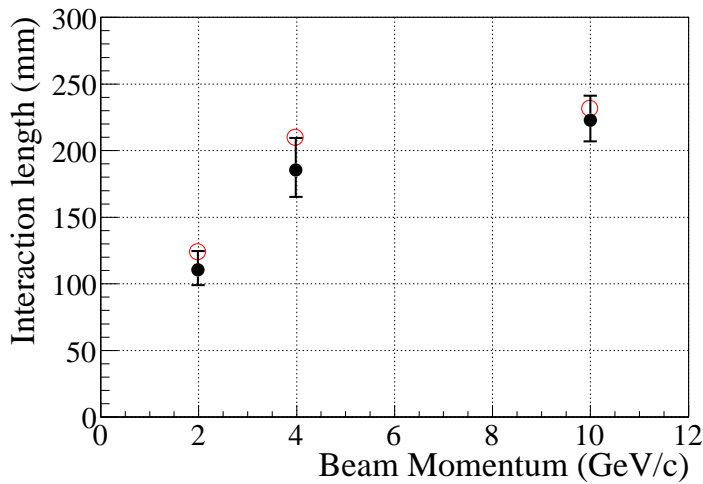}
		\caption[Interaction length as a function of beam momentum]
{Interaction length as a function of beam momentum.
Black dots with error bars (outlined circles) show experimental data (simulated data).}
		\label{fig:figure_6}
	\end{center}
\end{figure}

In total, 318 interactions (77 for 2~GeV/$c$, 68 for 4~GeV/$c$, 173 for 10~GeV/$c$) have been
found from the interaction measurements. Out of the found interactions of 2 GeV/$c$ (10 GeV/$c$), 
one (three) took place in the base of a film and one (one) in an emulsion layer of a film.
All the others occurred in the lead plates. 
The statistics of the interaction measurements are summarized in Table~\ref{table:beam_stop_results} and 
evaluated interaction lengths are also presented.
The interaction length, $\lambda $, is calculated as follows:
\begin{eqnarray}
	\lambda = - \frac{L}{{\rm ln}(1 - \frac{N}{N_0})}
\end{eqnarray}
where $L$ is the thickness of a lead plate and an emulsion film ($L = 1293$~\textmu m), 
$N$ is the number of found interactions and $N_0$ is the sum of the numbers of
followed tracks in all analyzed films.
Figure \ref{fig:figure_6} shows momentum dependence of the interaction length, 
which is in good agreement with the Monte Carlo simulation within the statistical errors.

\begin{table}[t]
	\caption[Results of the interaction measurements]
{Results of the interaction measurements and of the simulated data (MC) analysis.
Number of tracks followed, total track length followed in the ECC brick, 
number of interactions in the ECC brick, evaluated interaction length 
for each momentum beam are presented.}
	\label{table:beam_stop_results}
		\begin{center}
		\begin{tabular}{ccccccc} \hline \hline

			$P$ [GeV/$c$] & 2 & 2 (MC) & 4 & 4 (MC) & 10 & 10 (MC) \\ \hline
			Tracks & 584 & 11301 & 913 & 9260 & 2205 & 13746 \\
			Total $L$ [mm] & 8506 & 191794 & 12620 & 162009 & 38534 & 240922 \\
			Interactions & 77 & 1544 & 68 & 773 & 173 & 1040 \\
			$\lambda$ [mm] & $109.8^{+14.1}_{-11.4}$ & $123.6^{+3.3}_{-3.1}$ & $184.9^{+24.2}_{-20.1}$
 & $208.9^{+7.8}_{-7.3}$ & $222.5^{+18.4}_{-15.8}$ & $231.0^{+7.4}_{-7.0}$ \\ \hline \hline
		\end{tabular}
	\end{center}
\end{table}

\subsection{Topological characteristics}
The multiplicity and kink angle distributions of relativistic charged secondary particles 
are shown in Figure~\ref{fig:figure_7}. Topological characteristics are summarized 
in Table~\ref{table:Topological characteristics} and they are compared with those of simulated data.
As the beam momentum increases, the average number of secondary particles $<n>$ becomes larger.
In the case of inelastic interactions, the higher is the momentum, the smaller is the emission angle.
This tendency can be seen for 4~GeV/$c$ and 10~GeV/$c$ interactions, 
while it is not so for 2~GeV/$c$ interactions and others.
This might be explained by the fact that fraction of elastic interactions is larger 
in 2~GeV/$c$ interactions~\citep{pdg}.
For the average multiplicity $<n>$, the agreement between the experimental data and 
the simulated data is not so good in 2~GeV/$c$ interactions.
The other experimental data agree well with the simulation data.

\begin{table}[t]
	\caption[Topological characteristics]{Topological characteristics of experimental data and 
simulated data (MC). Number of events, average charged particle multiplicity $<n>$, 
number of 1-prong events, 
average kink angle for 1-prong events $<\theta_{\rm kink}>$,
number of 3-prong events,
average of kink angle average for 3-prong events $<\overline{\theta_{\rm kink}}>$
for 2, 4, 10 GeV/$c$ $\pi^-$ interactions are summarized.}
	\label{table:Topological characteristics}
		\begin{center}
		\begin{tabular}{ccccccc} \hline \hline
			$P$~[GeV/$c$ ] & 2 & 2 (MC) & 4 & 4 (MC) & 10 & 10 (MC)\\ \hline
			Events & 77 & 1544 & 68 & 773 & 173 & 1040 \\
			$<n>$ & 0.48 & 0.65 & 0.93 & 1.03 & 2.45 & 2.41 \\ \hline \hline
			1-prong events & 33 & 915 & 29 & 372 & 26 & 216 \\
			$<\theta_{\rm kink}>$ [rad] & 0.13 & 0.13 & 0.23 & 0.21 & 0.20 & 0.15 \\ \hline
			3-prong events & 0 & 3 & 2 & 32 & 44 & 246 \\
			$<\overline{\theta_{\rm kink}}>$ [rad] & - & 0.36 & 0.32 & 0.30 & 0.27 & 0.24 \\ \hline \hline
		\end{tabular}
	\end{center}
\end{table}

{\setkeys{Gin}{width=\subfigwidth}
\def\subfigtopskip{2pt}     
\def\subfigbottomskip{-5pt} 
\def\subfigcapskip{-5pt}    
\begin{figure}[htbp]
  \setlength{\subfigwidth}{.34\linewidth}
  \addtolength{\subfigwidth}{-.3\subfigcolsep}
  \begin{minipage}[b]{\subfigwidth}
    \subfigure{\includegraphics[width=1.05\columnwidth ]{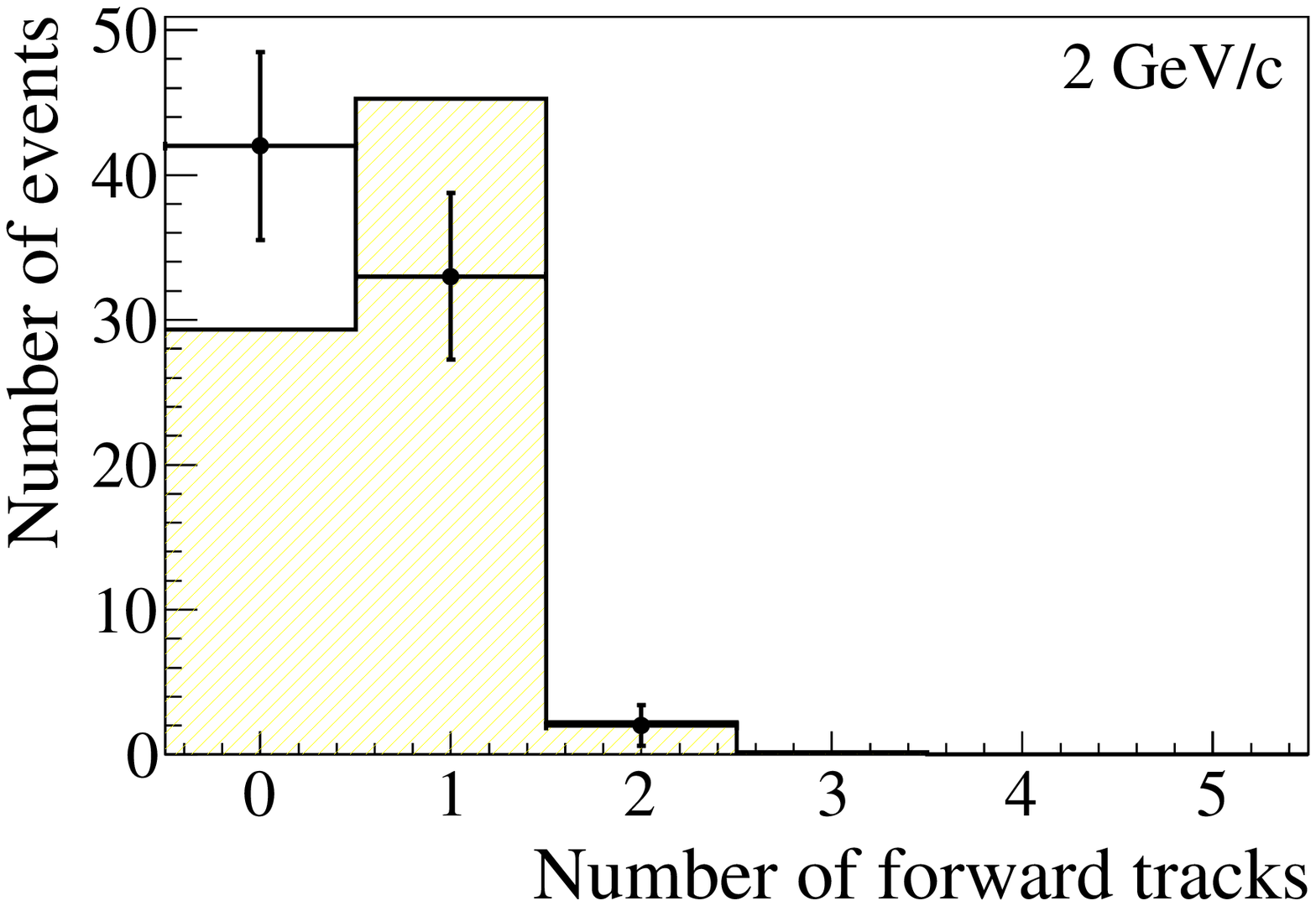}
   \label{fig:2gev_nt}}
  \end{minipage}
  \begin{minipage}[b]{\subfigwidth}
    \subfigure{\includegraphics[width=1.05\columnwidth ]{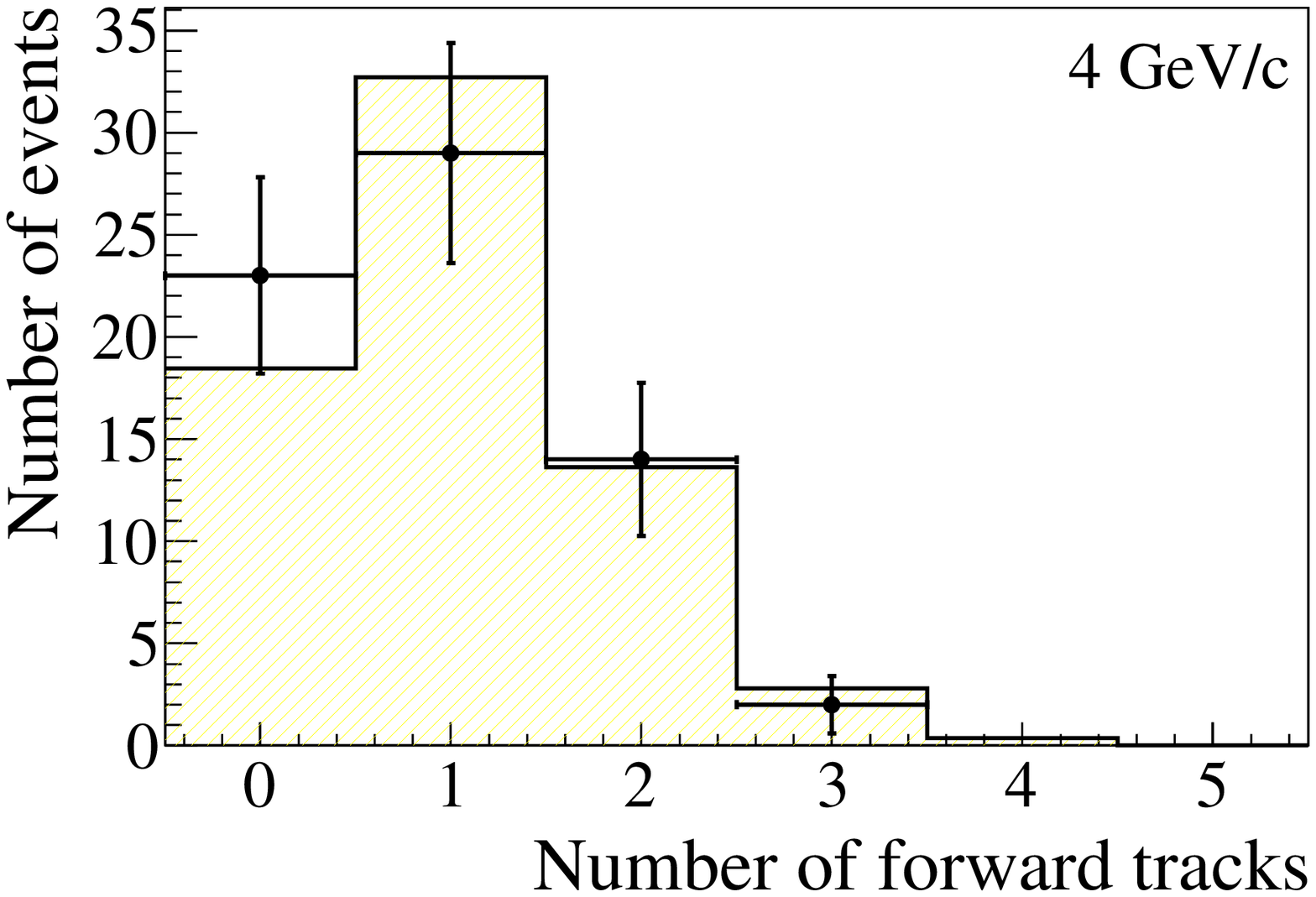}
   \label{fig:4gev_nt}}
  \end{minipage}
  \begin{minipage}[b]{\subfigwidth}
    \subfigure{\includegraphics[width=1.05\columnwidth ]{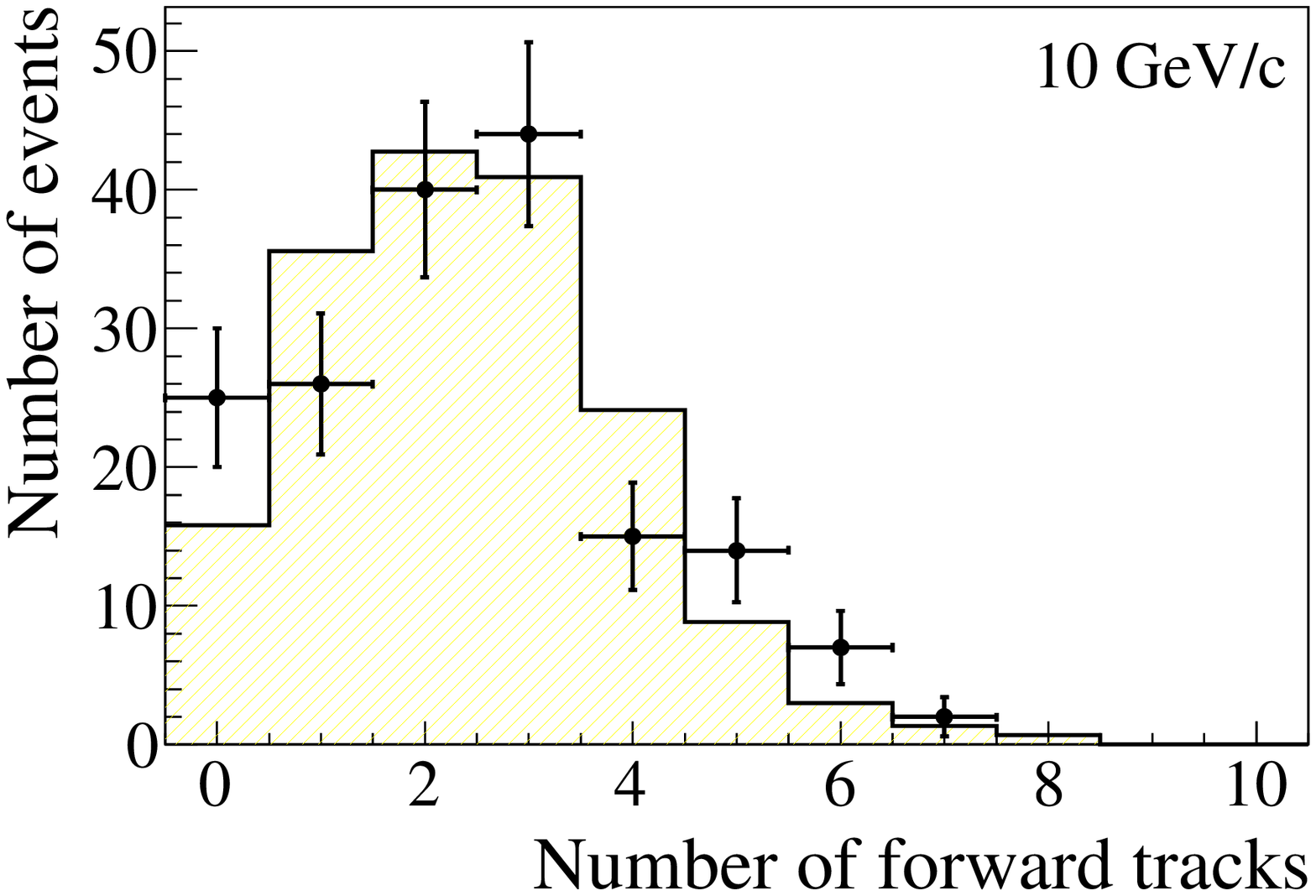}
   \label{fig:10gev_nt}}
  \end{minipage}
  \begin{minipage}[b]{\subfigwidth}
    \subfigure{\includegraphics[width=1.05\columnwidth ]{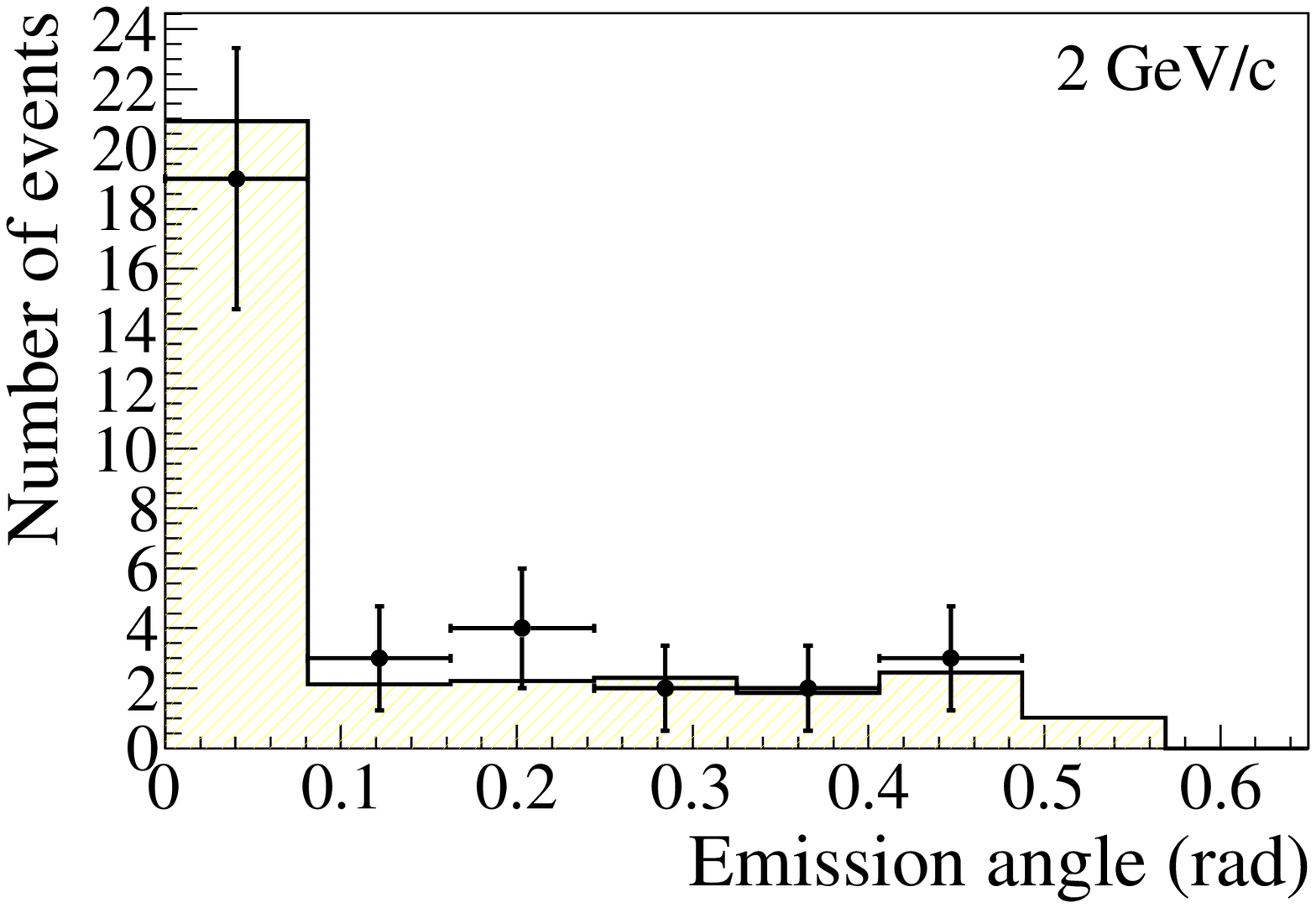}
   \label{fig:2gev_kink_1prong}}
  \end{minipage}
  \begin{minipage}[b]{\subfigwidth}
    \subfigure{\includegraphics[width=1.05\columnwidth ]{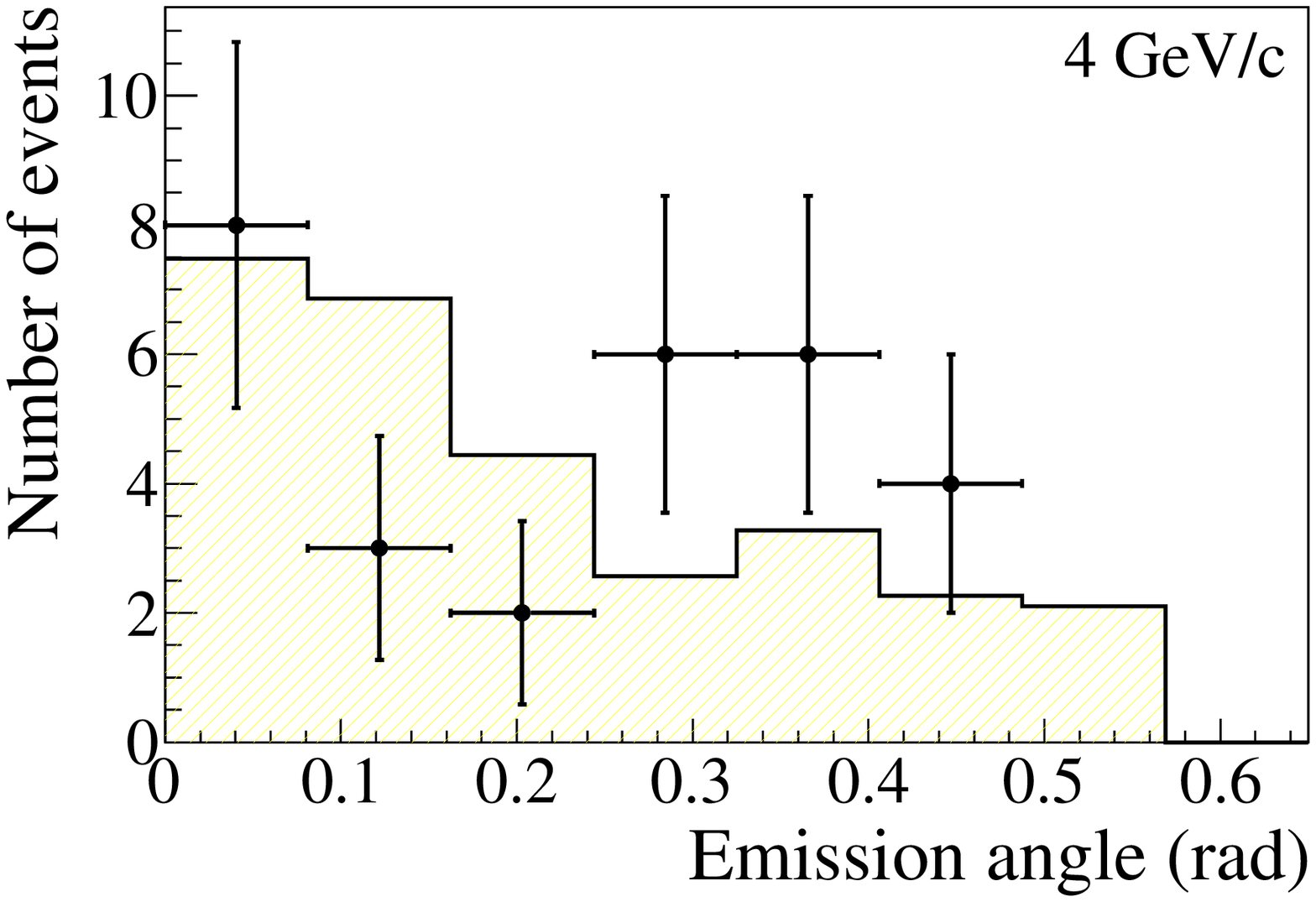}
   \label{fig:4gev_kink_1prong}}
  \end{minipage}
  \begin{minipage}[b]{\subfigwidth}
    \subfigure{\includegraphics[width=1.05\columnwidth ]{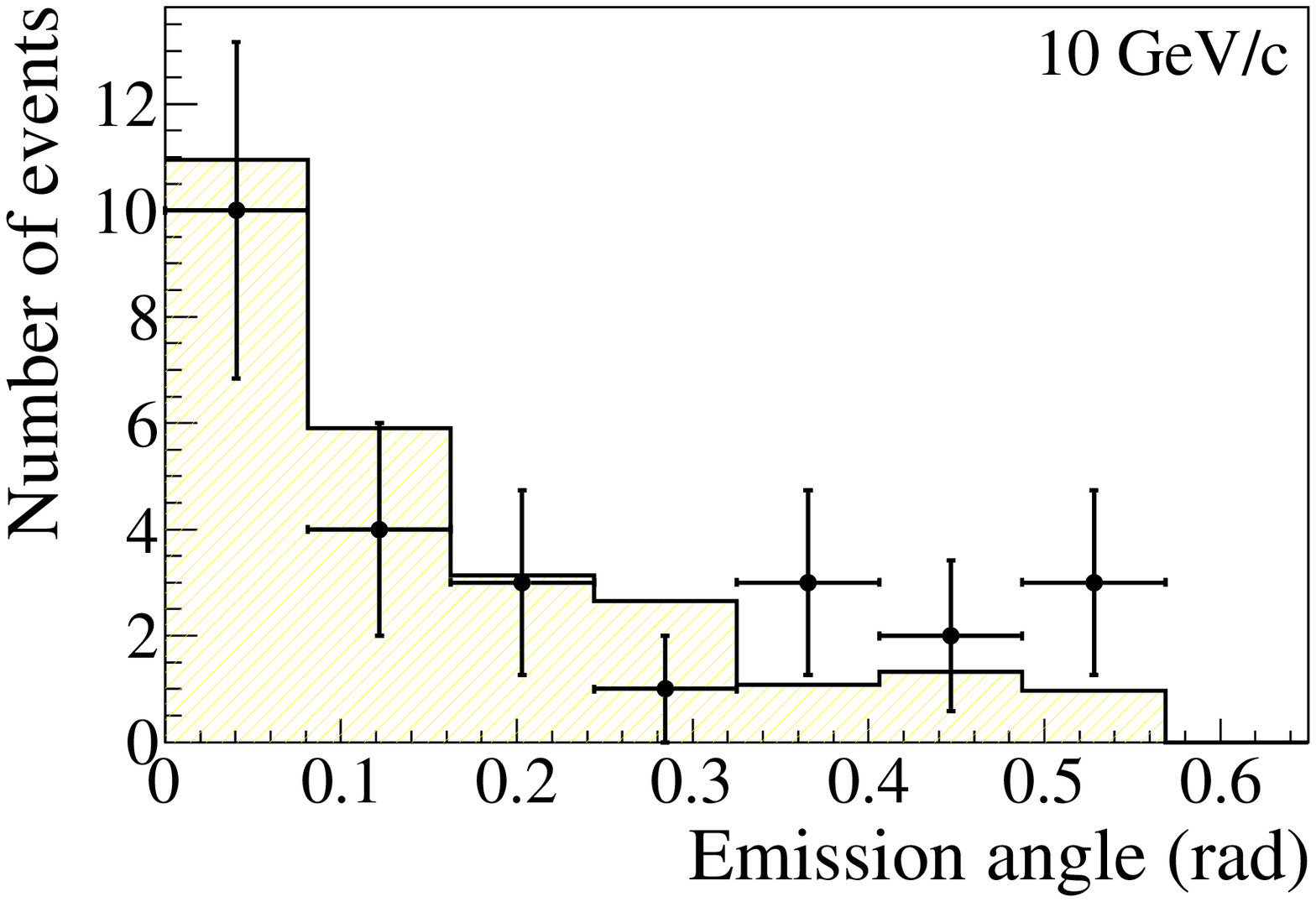}
   \label{fig:10gev_kink_1prong}}
  \end{minipage}
  \caption{Multiplicity distributions of relativistic particle tracks in the forward hemisphere (Top)
and kink angle distributions of 1-prong events (Bottom)
for experimental data (dots with error bars) and simulated data (histogram).
The simulated distributions are normalized to the real data.}
  \label{fig:figure_7}
\end{figure}
}

\subsection{Nuclear fragment association}

\begin{figure}[htbp]
	\begin{center}
		\includegraphics[height=7cm]{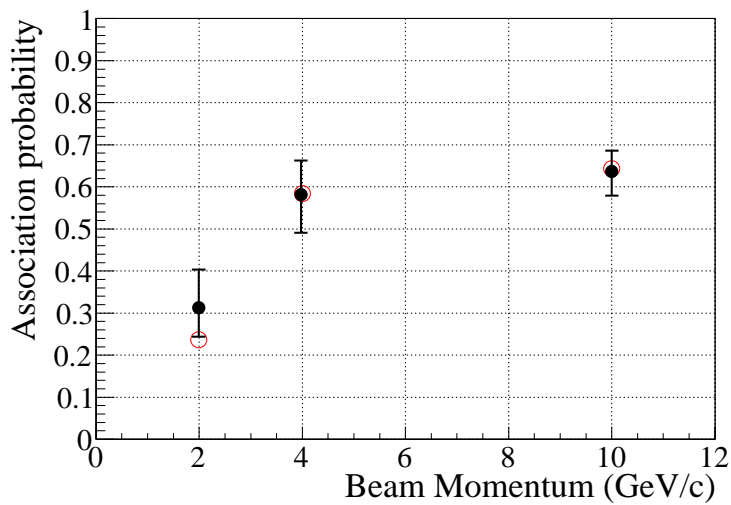}
		\caption[Nuclear fragments associated probability]
{Association probability of nuclear fragments as a function of beam momentum.
Black dots with error bars (outlined circles) show experimental data (simulated data).}
		\label{fig:figure_8}
	\end{center}
\end{figure}

We present measurement results of nuclear fragments for 1-prong and 3-prong events,
which are relevant to topologies of the $\tau$ hadronic decay, 
and corresponding results for simulated data in Table~\ref{table:black_results_Data&MC}.
Here numbers of events associated with at least one heavily ionizing particle,
pointing to the vertex, found in either the upstream film or the downstream film of the vertex
are listed.
The probability is the fraction of fragment associated events in the 1-prong or 3-prong events.
Figure~\ref{fig:figure_8} shows the association probability as a function of the beam momentum, where
data of 1-prong and 3-prong events are merged, together with simulation results.
At least one nuclear fragment particle is associated with about a half of events
for beam momentum greater than 4~GeV/$c$.
The experimental data of the fragment association probability 
are well reproduced by the simulation with differences less than 10\%.
The number of background nuclear fragment tracks was estimated to be 0.035 tracks/event
by scanning around dummy vertices with the same conditions.
Figure~\ref{fig:figure_9} shows good agreement in multiplicity and polar angle distributions
of nuclear fragments between experimental data (dots with error bars) and simulated data
(histogram).

\begin{table}[t]
	\caption[Nuclear fragment search results]{Results of nuclear fragment search 
(experimental data and simulated data). Lower limit of the probability shows the value
at 90\% confidence level.}
	\label{table:black_results_Data&MC}
		\begin{center}
		\begin{tabular}{ccccccc} \hline \hline

			$P$~[GeV/$c$ ] & \multicolumn{2}{c}{2} & \multicolumn{2}{c}{4} & \multicolumn{2}{c}{10} \\ \hline
			Prong & 1 & 3 & 1 & 3 & 1 & 3 \\ \hline 
			Events & 32 & 0 & 29 & 2 & 25 & 41 \\
			Fragment associated & 10 & 0 & 16 & 2 & 15 & 27 \\
			Probability [\%] & $31.3 ^{+9.1}_{-6.9}$ & - & $55.2^{+8.6}_{-9.3}$ &
 $ > 46.5$ & $60.0 ^{+8.9}_{-10.2}$ & $65.9 ^{+6.5}_{-8.0}$ \\ \hline \hline

			Events (MC) & 908 & 0 & 372 & 32 & 214 & 246 \\
			Fragment associated (MC) & 213 & 0 & 219 & 17 & 127 & 169 \\
			Probability (MC) [\%] & $23.5 ^{+1.5}_{-1.3}$ & - & $58.9^{+2.5}_{-2.6}$ &
 $53.1 ^{+8.4}_{-8.8}$ & $59.3 ^{+3.3}_{-3.5}$ & $68.7 ^{+2.8}_{-3.1}$ \\ \hline \hline
		\end{tabular}
	\end{center}
\end{table}

{\setkeys{Gin}{width=\subfigwidth}
\def\subfigtopskip{2pt}     
\def\subfigbottomskip{-5pt} 
\def\subfigcapskip{-5pt}    
\begin{figure}[htbp]
  \setlength{\subfigwidth}{.34\linewidth}
  \addtolength{\subfigwidth}{-.3\subfigcolsep}
  \begin{minipage}[b]{\subfigwidth}
    \subfigure{\includegraphics[width=1.05\columnwidth ]{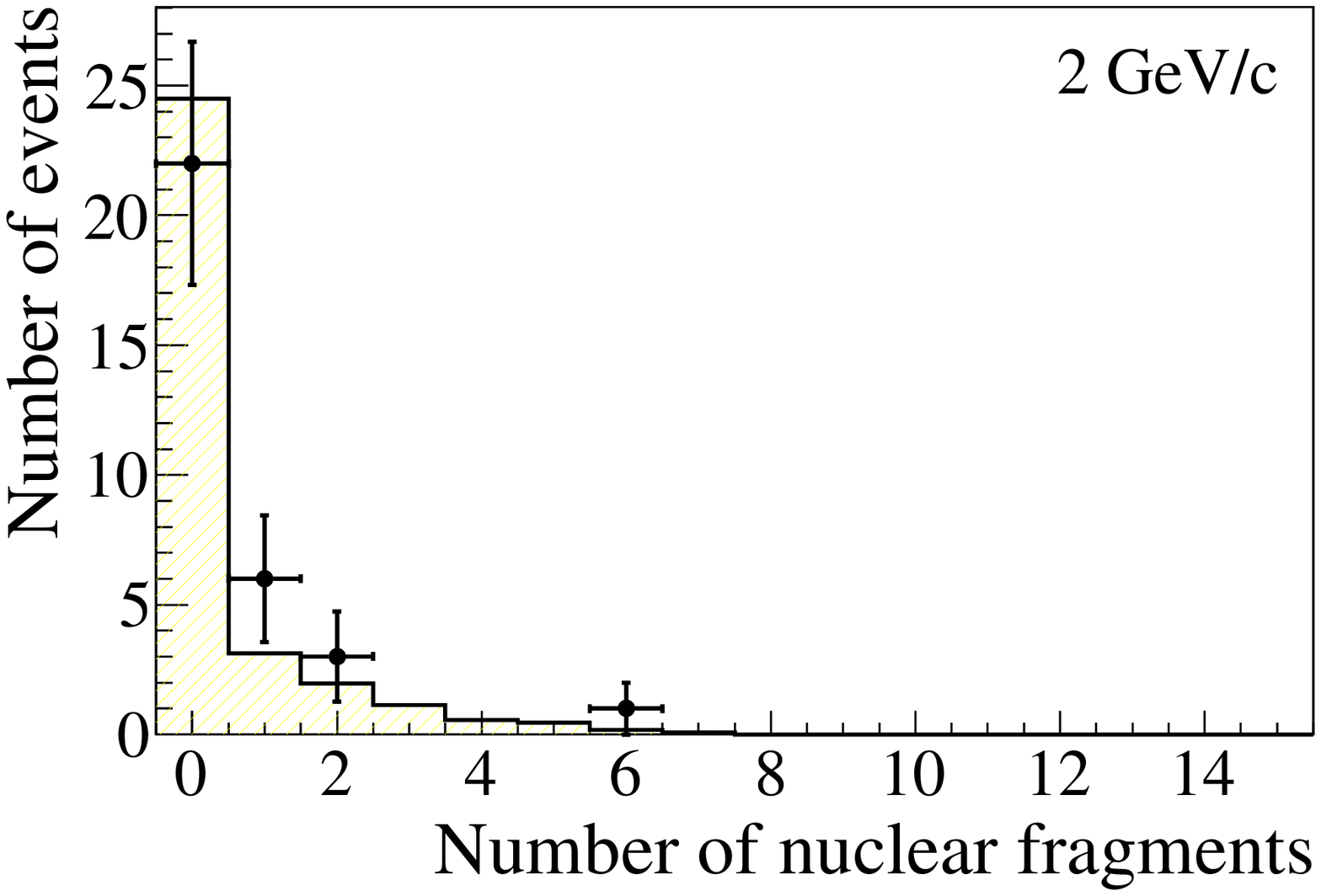}
   \label{fig:2gev_black_multi}}
  \end{minipage}
  \begin{minipage}[b]{\subfigwidth}
    \subfigure{\includegraphics[width=1.05\columnwidth ]{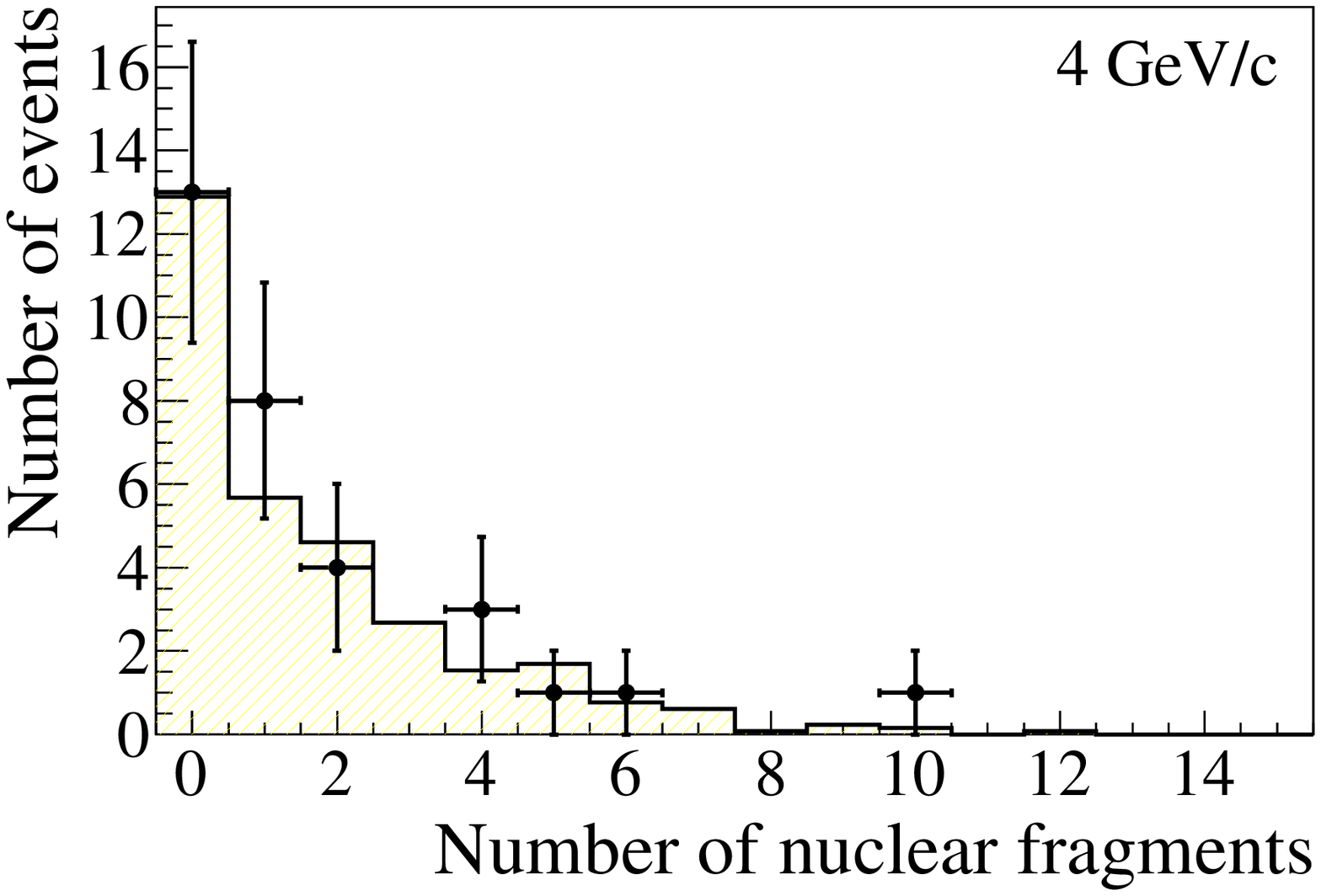}
   \label{fig:4gev_black_multi}}
  \end{minipage}
  \begin{minipage}[b]{\subfigwidth}
    \subfigure{\includegraphics[width=1.05\columnwidth ]{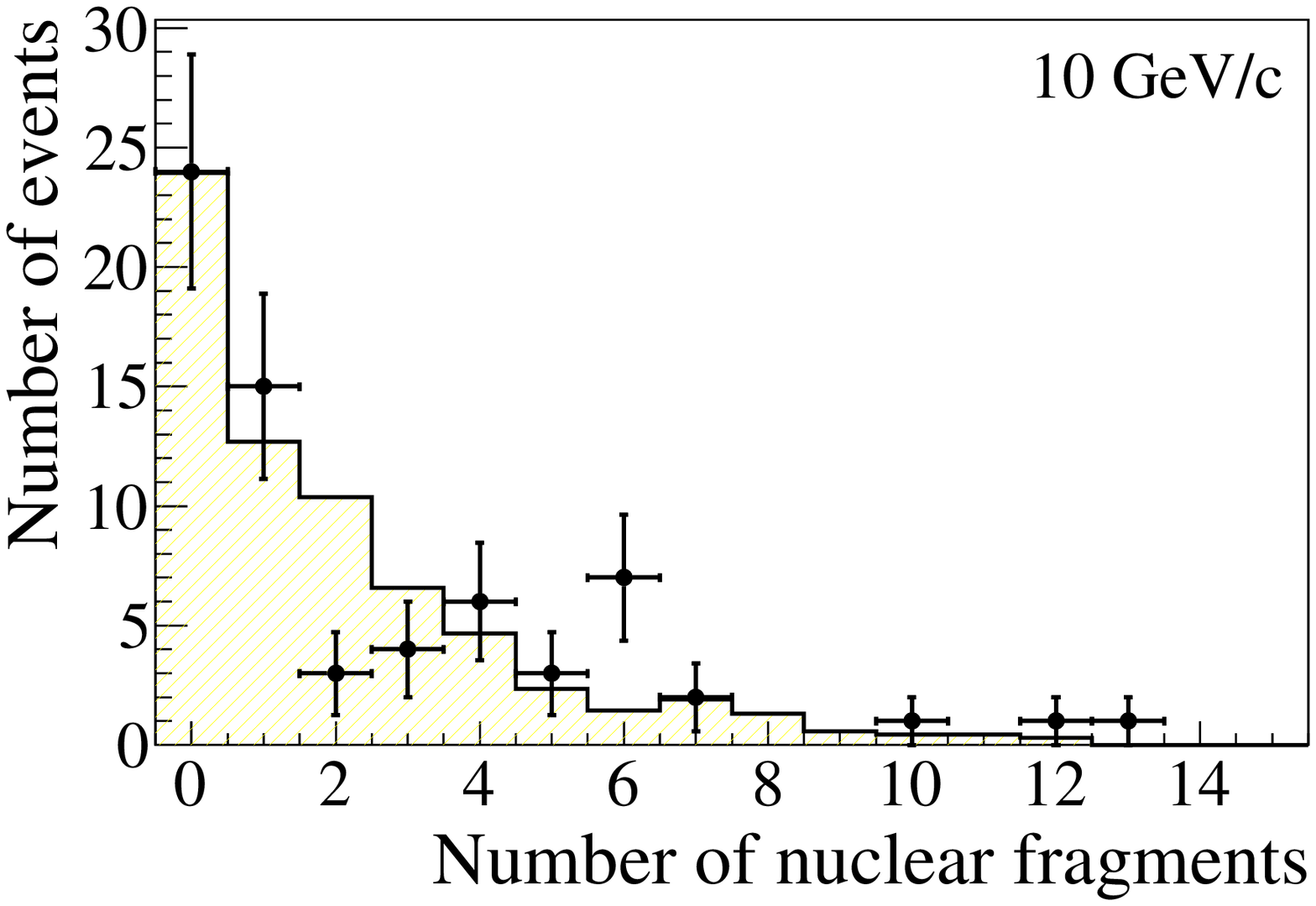}
   \label{fig:10gev_black_multi}}
  \end{minipage}
  \begin{minipage}[b]{\subfigwidth}
    \subfigure{\includegraphics[width=1.05\columnwidth ]{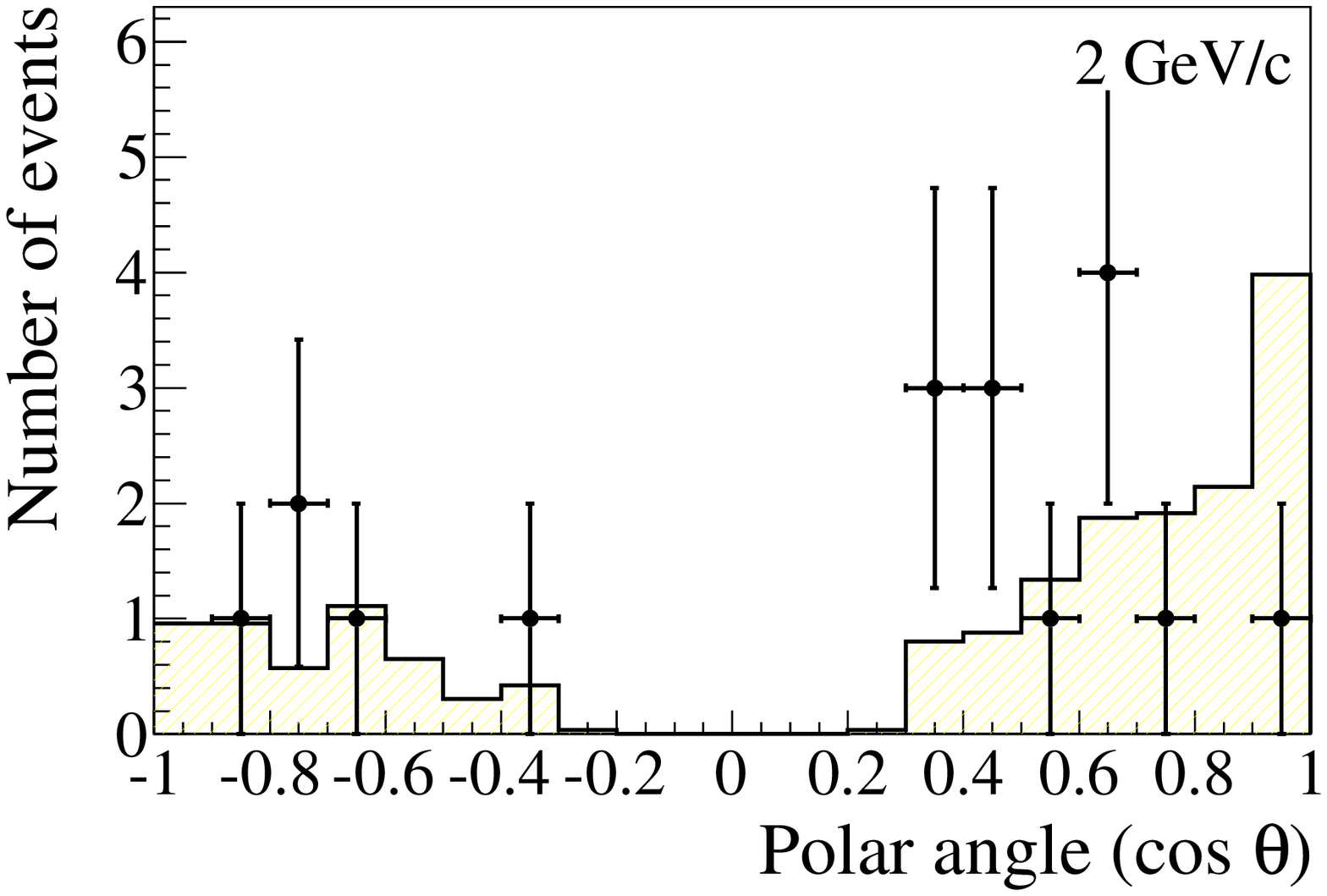}
   \label{fig:2gev_black_cos}}
  \end{minipage}
  \begin{minipage}[b]{\subfigwidth}
    \subfigure{\includegraphics[width=1.05\columnwidth ]{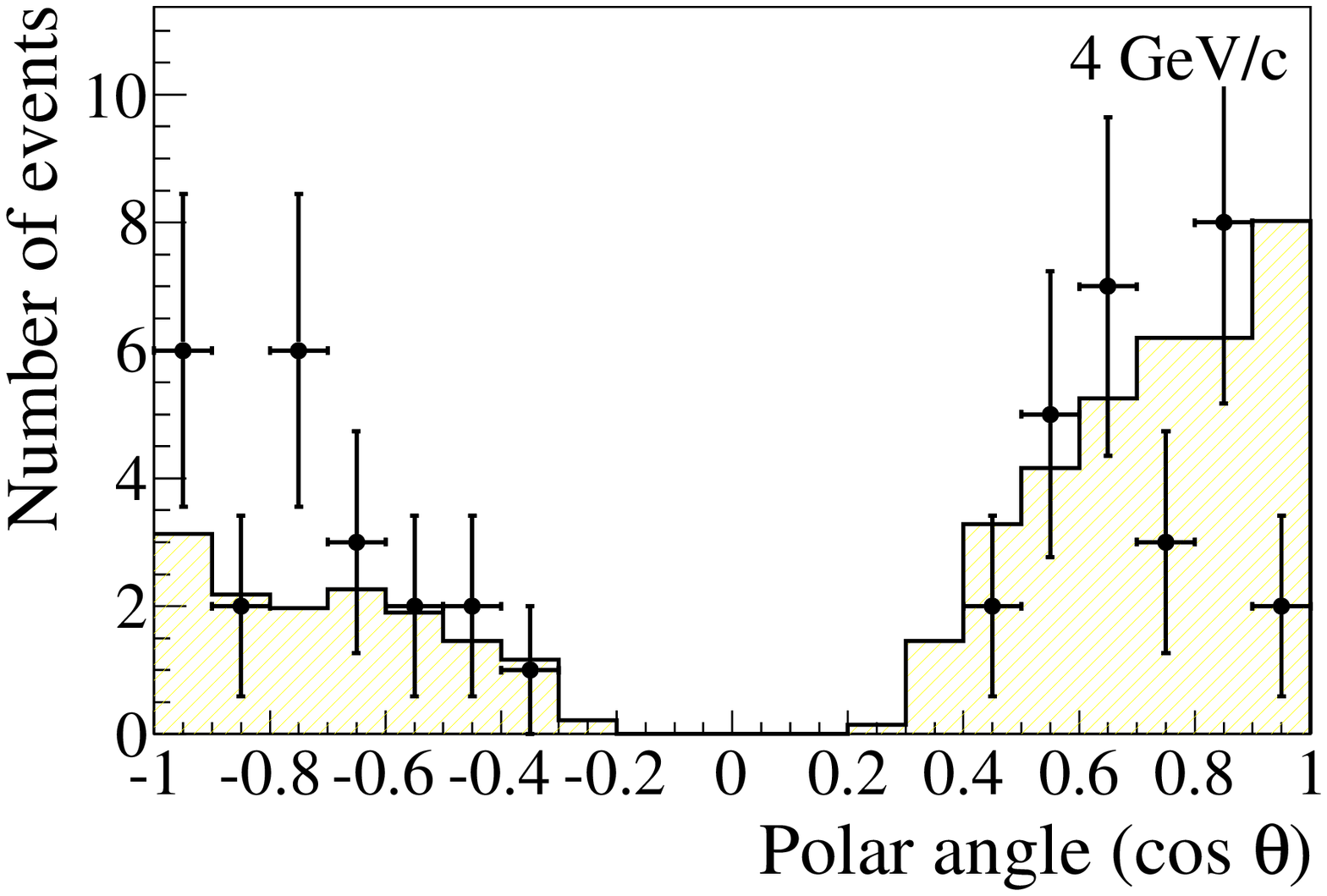}
   \label{fig:4gev_black_cos}}
  \end{minipage}
  \begin{minipage}[b]{\subfigwidth}
    \subfigure{\includegraphics[width=1.05\columnwidth ]{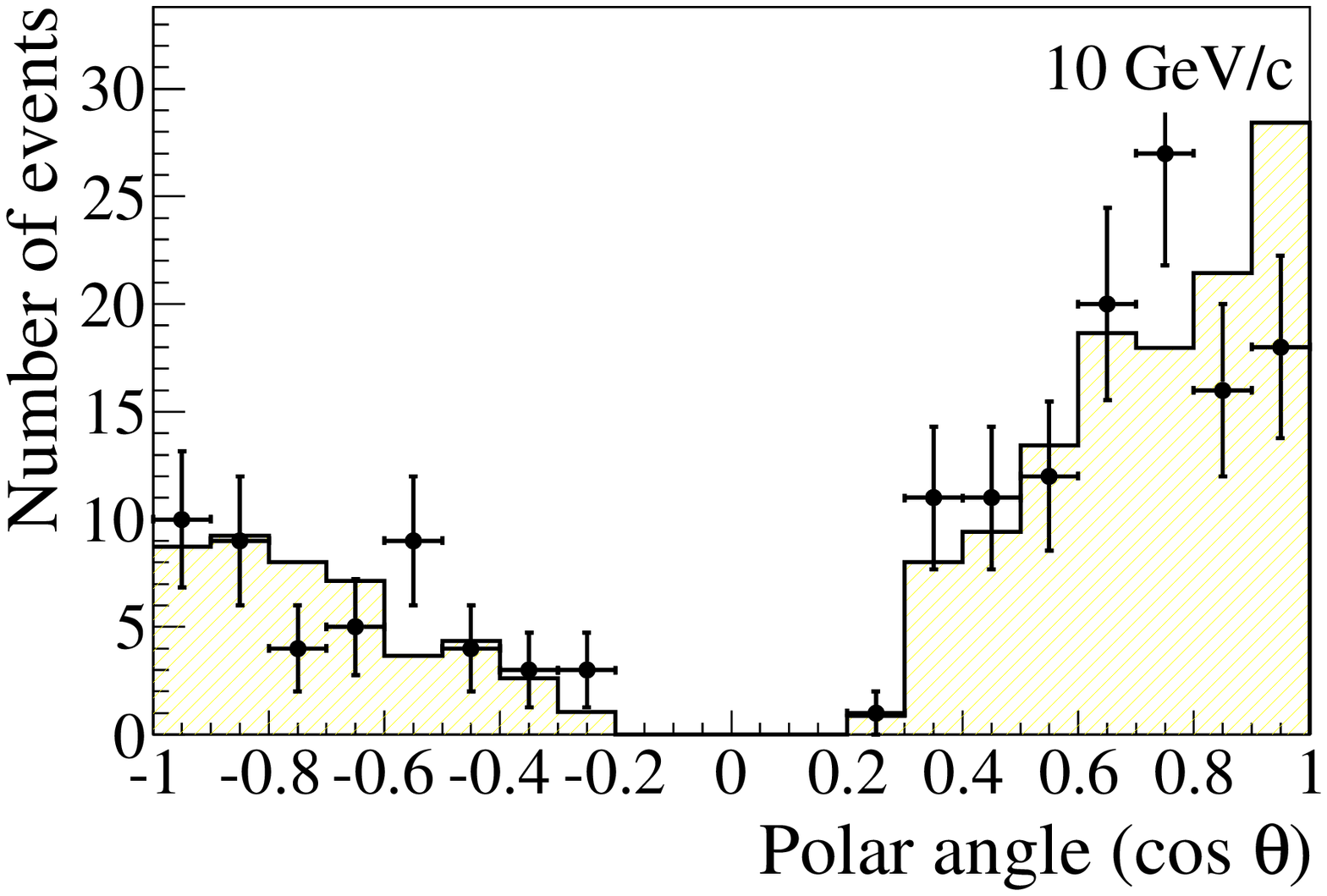}
   \label{fig:10gev_black_cos}}
  \end{minipage}
  \caption{Top: multiplicity distributions of nuclear fragments 
for experimental data (dots with error bars) and simulated data (histogram).
Bottom: polar angle distributions of nuclear fragments 
for experimental data (dots with error bars) and simulated data (histogram).}
  \label{fig:figure_9}
\end{figure}
}

\subsection{Kinematical characteristics}

In OPERA, for $\tau$ candidates decaying into a single charged hadron, kinematical selection cuts; 
$p > 2$~GeV/$c$ and $p_{\rm T} > 0.6$~GeV/$c$
are required~\citep{first_tau,opera_2012status}~\footnote{
The cut on $p_{\rm T}$ is 0.3~GeV/$c$ if there is a converted $\gamma$ attached
to the secondary vertex. In this analysis, no attempt is made to find $\gamma$s
because the numbers of lead plates and emulsion films are limited.}.
Scatter plots of $p_{\rm T}$ versus $p$ of the secondary particles
of 1-prong hadron interactions are shown together with those of simulated data
in Figure~\ref{fig:figure_10}.
For comparison, scatter plots of $p_{\rm T}$ versus $p$ of the secondary particles
of 10~GeV/$c$ 3-prong interactions are also shown in Figure~\ref{fig:figure_11}.
Statistics are limited because 
only secondary tracks for which more than 14 lead plates are available as scattering materials
are able to be momentum-reconstructed.
In this condition, error of the momentum measurement was estimated to be $\sigma _p = 45 (55) \% $ 
at $p \leq 4~{\rm GeV}/c \;(p > 4~{\rm GeV}/c)$ 
by using the multiple scattering measurements of beam pions with known momentum, 
i.e. 2, 4, 10 GeV/$c$, under the same condition.
In Figure~\ref{fig:figure_10}, the dark area defines the domain in which $\tau $ 
decay candidates are selected and the hatched area defines the domain rejected 
by the selection cuts ($\theta _{\mbox{kink}} > 0.02$~rad) and S-UTS angle acceptance ($\tan \theta < 0.6$).
Table~\ref{table:selected tracks} summarizes the fraction of secondary particles
in the selection domain for the events in which secondary particle momenta are measurable.
Comparisons between experimental and simulated data show generally good agreement,
however the final statistics of the tracks in the selection domain are not sufficient 
to obtain quantitative results on how good the agreement is.
We will therefore consider an alternative method to estimate the goodness of the agreement
in the next subsection.

{\setkeys{Gin}{width=\subfigwidth}
\def\subfigtopskip{2pt}     
\def\subfigbottomskip{-5pt} 
\def\subfigcapskip{-5pt}    
\begin{figure}[htbp]
  \setlength{\subfigwidth}{.34\linewidth}
  \addtolength{\subfigwidth}{-.3\subfigcolsep}
  \begin{minipage}[b]{\subfigwidth}
    \subfigure{\includegraphics[width=1.05\columnwidth ]{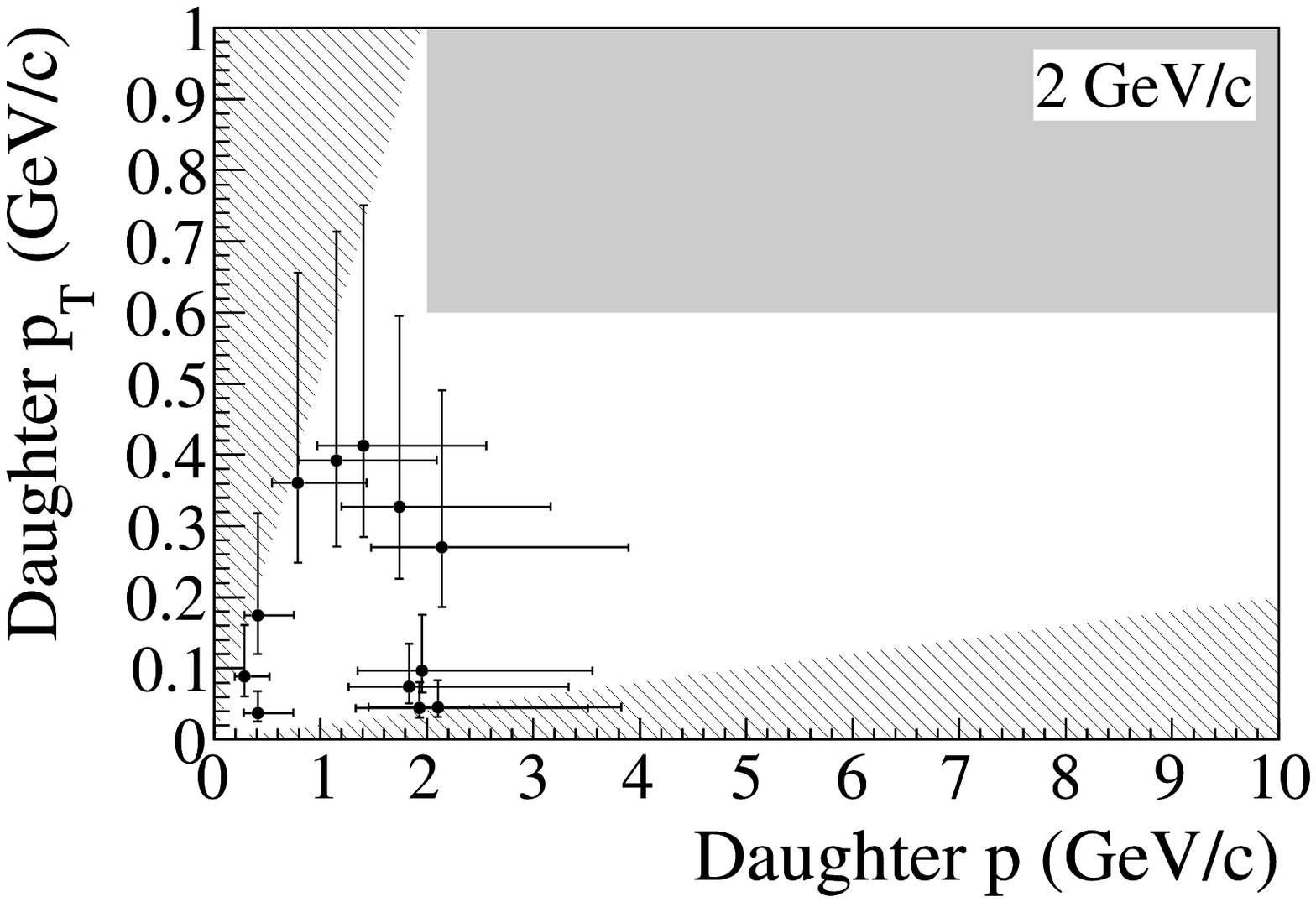}
   \label{fig:2gev_pt_1prong}}
  \end{minipage}
  \begin{minipage}[b]{\subfigwidth}
    \subfigure{\includegraphics[width=1.05\columnwidth ]{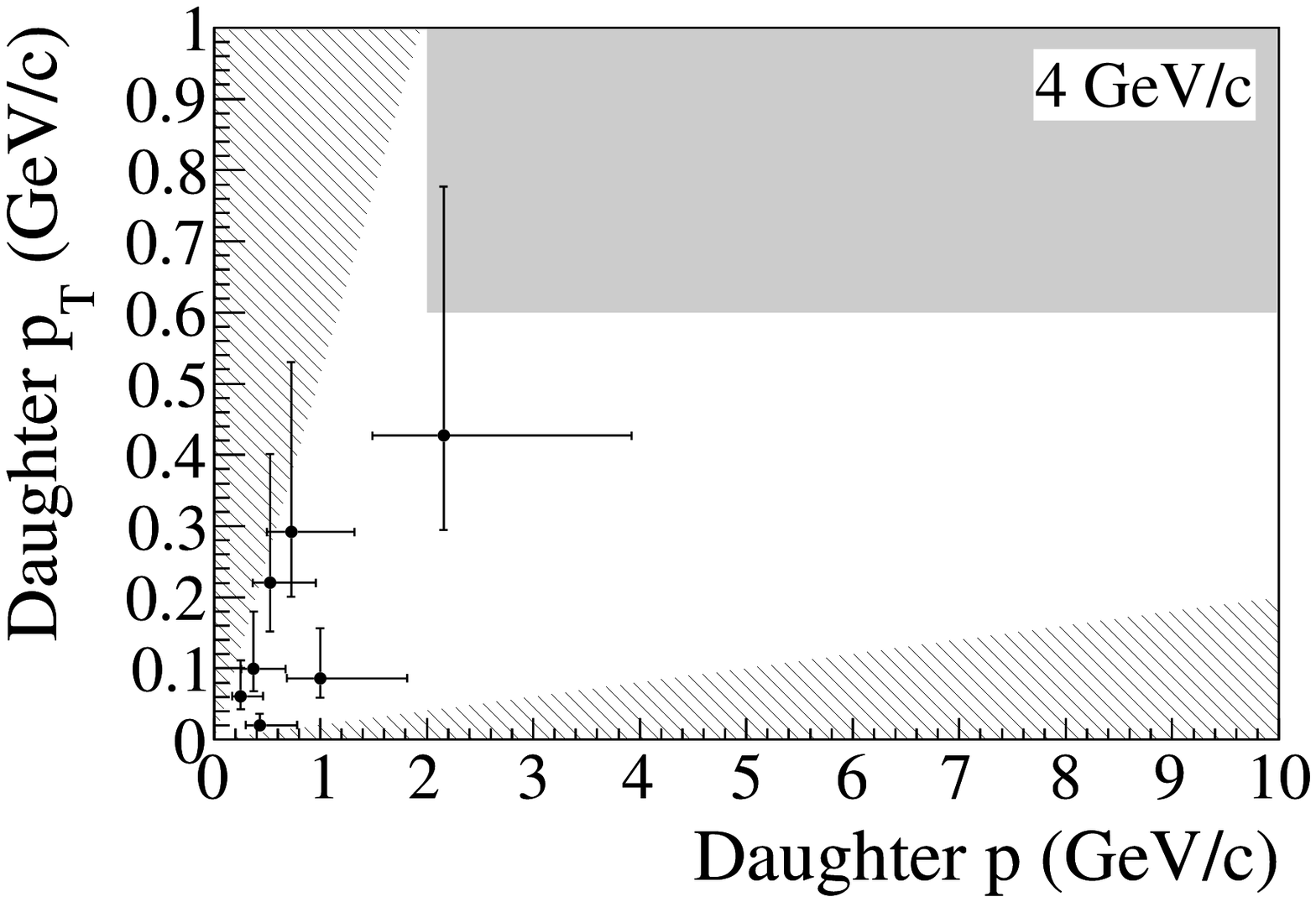}
   \label{fig:4gev_pt_1prong}}
  \end{minipage}
  \begin{minipage}[b]{\subfigwidth}
    \subfigure{\includegraphics[width=1.05\columnwidth ]{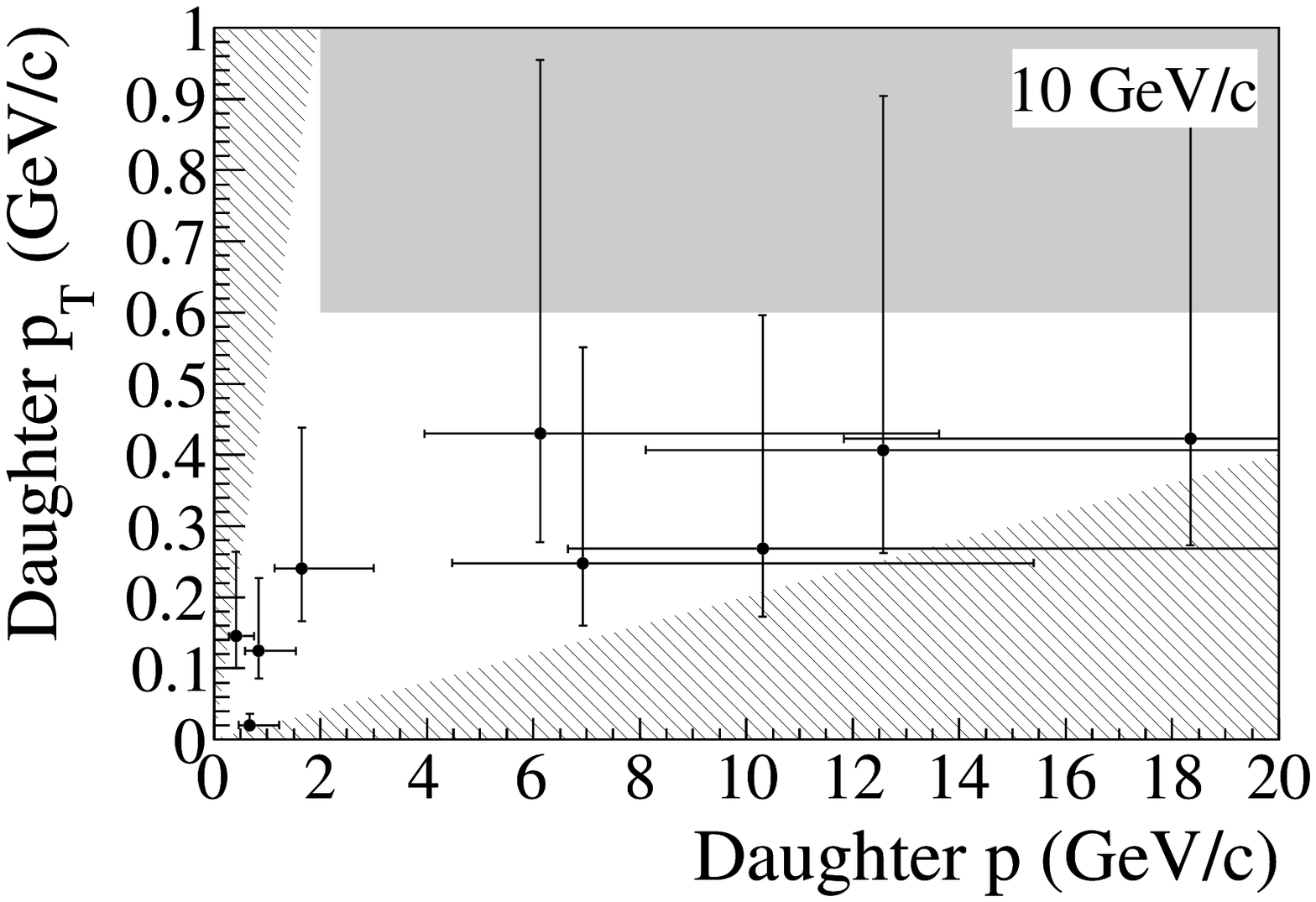}
   \label{fig:10gev_pt_1prong}}
  \end{minipage}
  \begin{minipage}[b]{\subfigwidth}
    \subfigure{\includegraphics[width=1.05\columnwidth ]{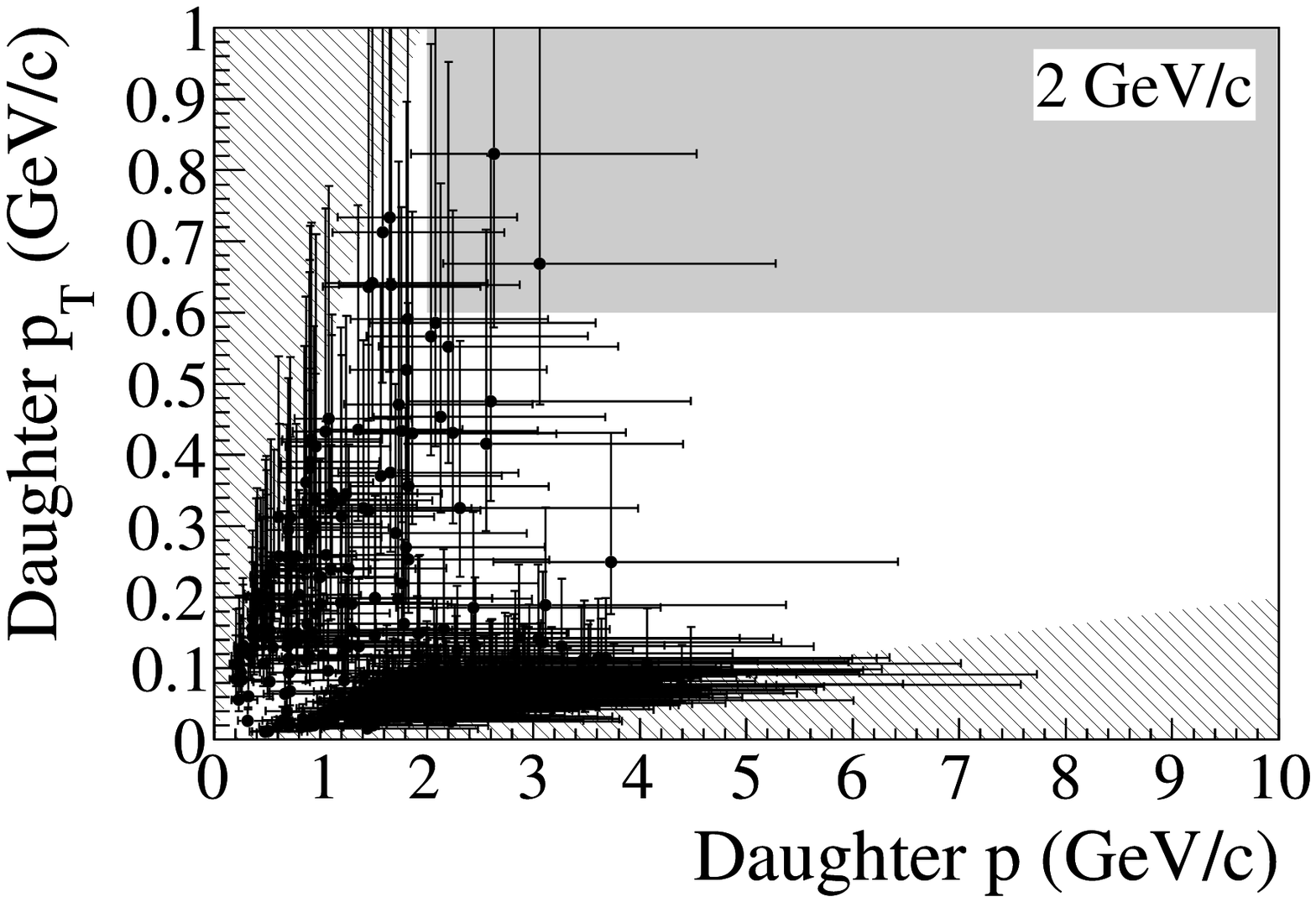}
   \label{fig:2gevmc_pt_1prong}}
  \end{minipage}
  \begin{minipage}[b]{\subfigwidth}
    \subfigure{\includegraphics[width=1.05\columnwidth ]{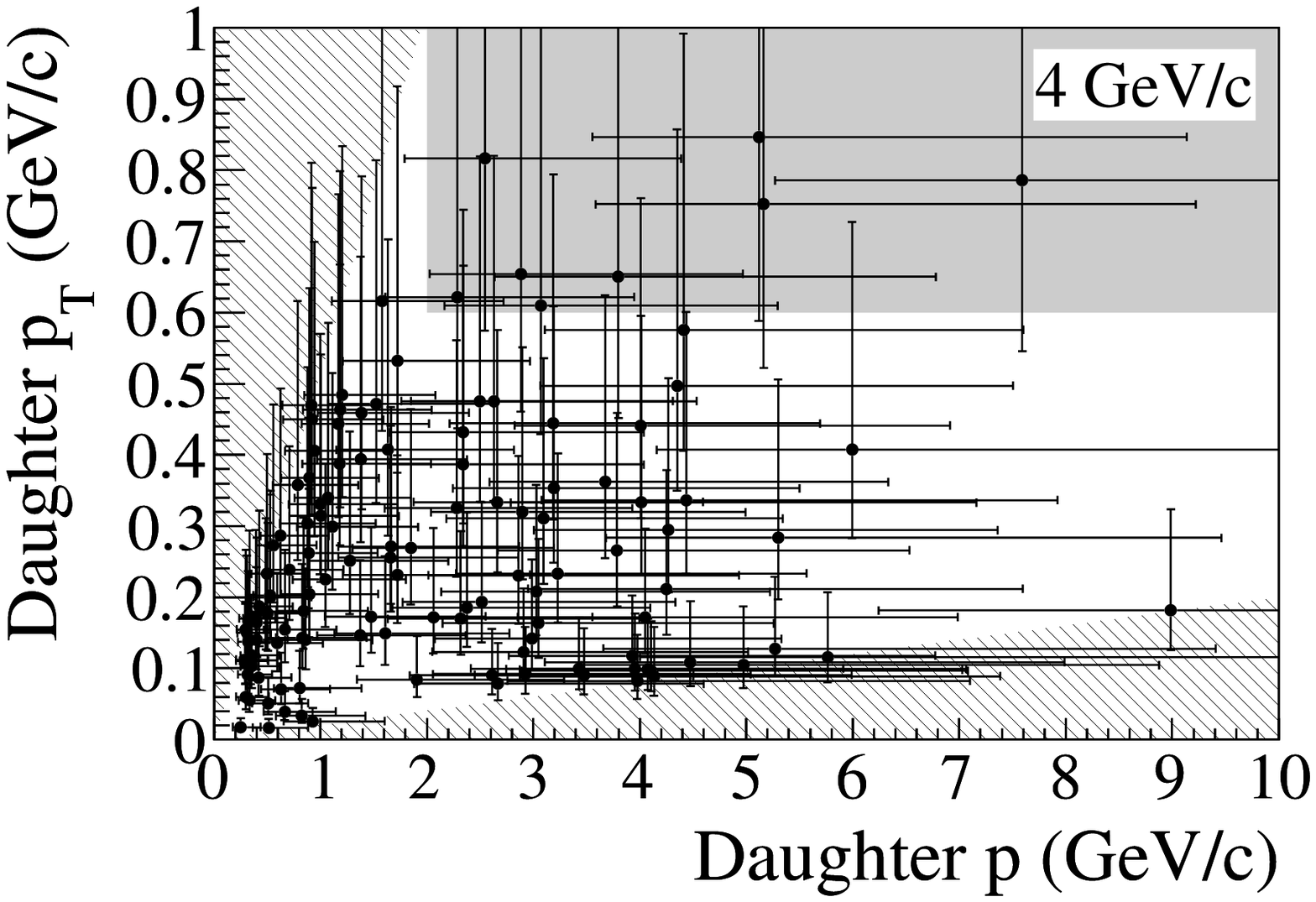}
   \label{fig:4gevmc_pt_1prong}}
  \end{minipage}
  \begin{minipage}[b]{\subfigwidth}
    \subfigure{\includegraphics[width=1.05\columnwidth ]{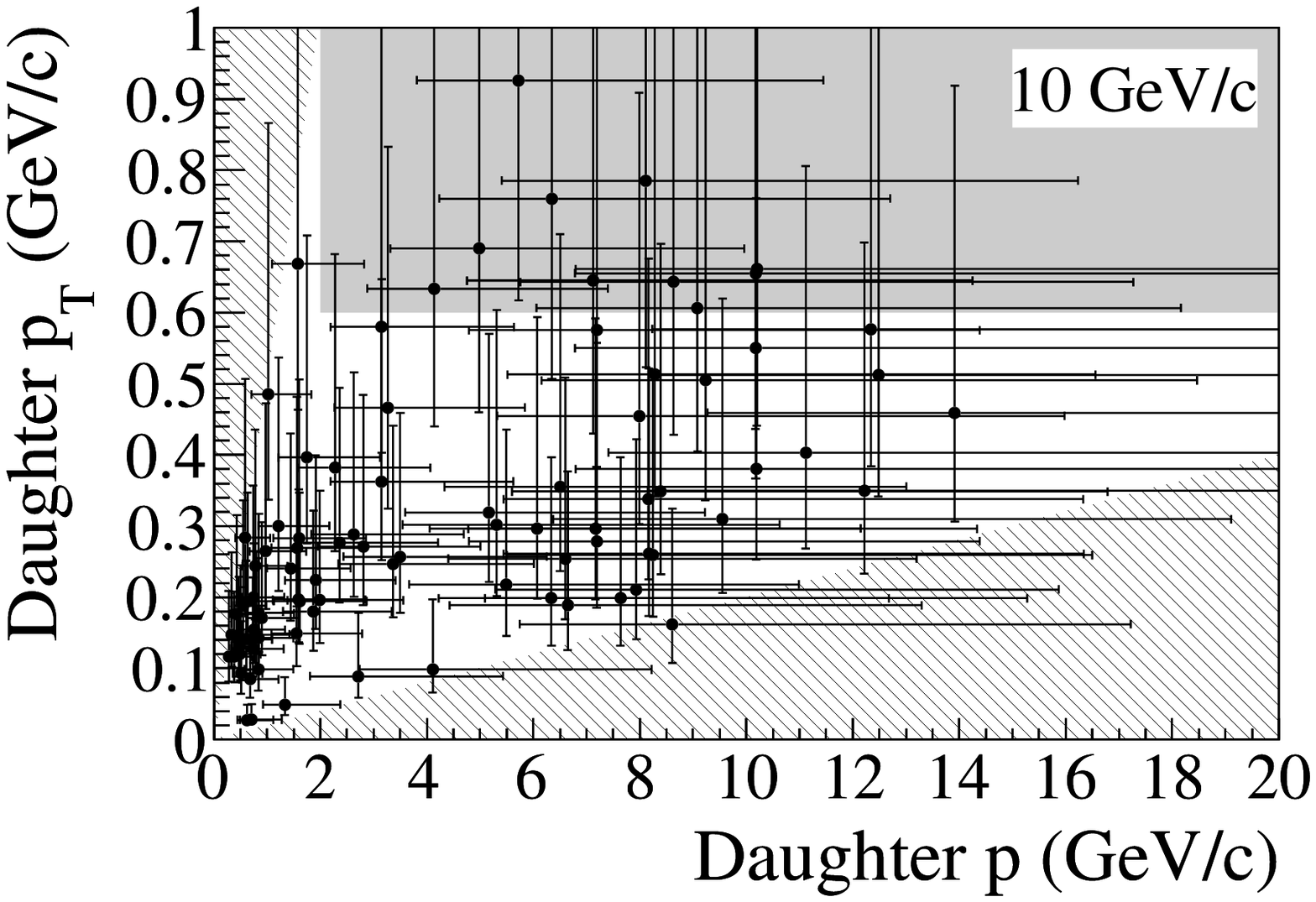}
   \label{fig:10gevmc_pt_1prong}}
  \end{minipage}
  \caption{Top: Scatter plot of $p_{\rm T}$ vs. $p$ of the secondary particle of 1-prong interactions (Experimental data).
  Bottom: Scatter plot of $p_{\rm T}$ vs. $p$ of the secondary particle of 1-prong interactions (Simulated data).
 On the figures the dark area defines the domain in which $\tau$ decay candidates are selected and
 the hatched area defines the domain rejected by the selection cuts.}
  \label{fig:figure_10}
\end{figure}
}

{\setkeys{Gin}{width=\subfigwidth}
\def\subfigtopskip{2pt}     
\def\subfigbottomskip{-5pt} 
\def\subfigcapskip{-5pt}    

\begin{figure}[hbp]
  \setlength{\subfigwidth}{.15\linewidth}
  \addtolength{\subfigwidth}{-.3\subfigcolsep}
  \begin{minipage}[b]{\subfigwidth}
	\hspace{-.3\subfigcolsep}
  \end{minipage}
  \setlength{\subfigwidth}{.34\linewidth}
  \addtolength{\subfigwidth}{-.3\subfigcolsep}
  \begin{minipage}[b]{\subfigwidth}
    \subfigure{\includegraphics[width=1.05\columnwidth ]{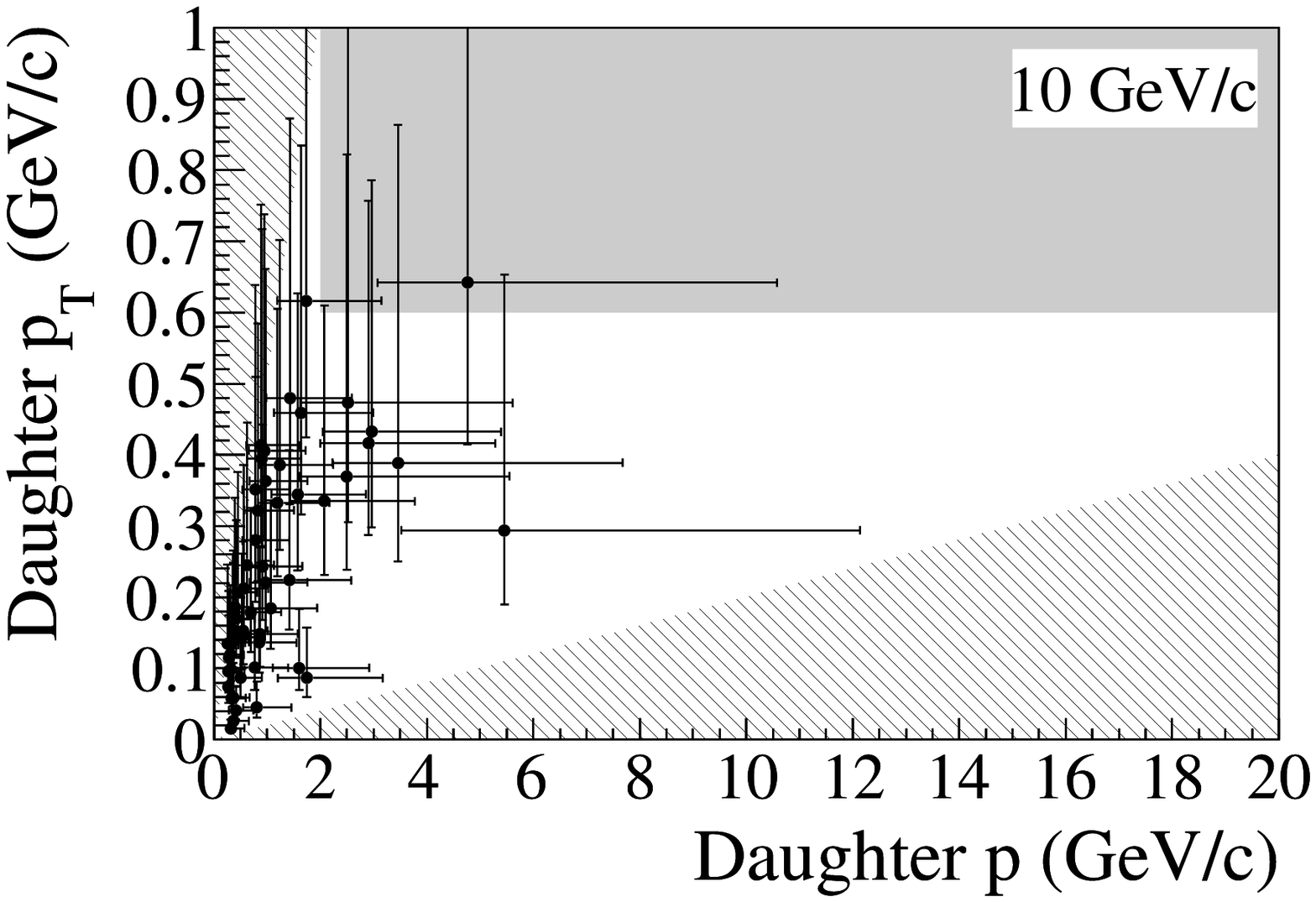}
   \label{fig:10gev_pt_3prong}}
  \end{minipage}
  \begin{minipage}[b]{\subfigwidth}
    \subfigure{\includegraphics[width=1.05\columnwidth ]{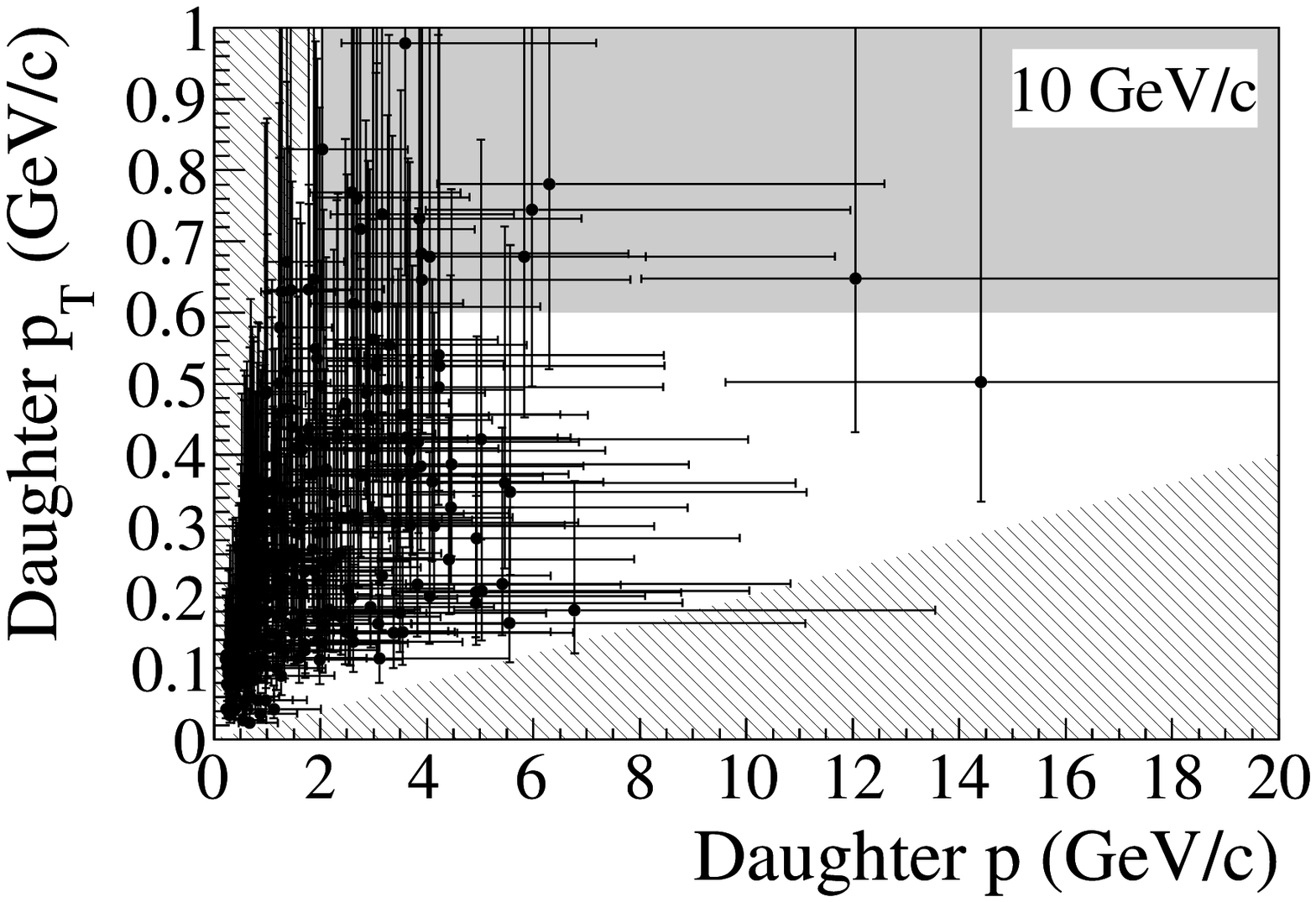}
   \label{fig:10gevmc_pt_3prong}}
  \end{minipage}
  \caption{Left: Scatter plot of $p_{\rm T}$ vs. $p$ of the secondary particle of 10 GeV/$c$ 3-prong interactions (Experimental data).
  Right: Scatter plot of $p_{\rm T}$ vs. $p$ of the secondary particle of 10 GeV/$c$ 3-prong interactions (Simulated data).
 On the figures the dark area defines the domain in which $\tau$ decay candidates are selected and
 the hatched area defines the domain rejected by the selection cuts.}
  \label{fig:figure_11}
\end{figure}
}

\begin{table}[t]	
	\caption[Fraction of secondary particles in the selection domain]
{Fraction of secondary particles in the selection domain of $p > 2$~GeV/$c$ and $p_{\rm T} > 0.6$~GeV/$c$.
Numbers of secondary tracks for which the momentum is measurable
and numbers of tracks in the selection domain are summarized.
Upper limits are at 90\% confidence level.}
	\label{table:selected tracks}
	\begin{center}
		\begin{tabular}{ccccccc} \hline \hline

			$P$~[GeV/$c$ ] & \multicolumn{2}{c}{2} & \multicolumn{2}{c}{4} & \multicolumn{2}{c}{10} \\ \hline
			Prong & 1 & 3 & 1 & 3 & 1 & 3 \\ \hline 
			Tracks $p$ measurable & 12 & 0 & 7 & 0 & 10 & 54 \\
			Tracks in selection domain & 0 & 0 & 0 & 0 & 1 & 1 \\
			Fraction [\%] & $< 16.3$ & - & $< 25.0$ & -
  & $10.0 ^{+16.0}_{-3.0}$ & $1.9^{+3.7}_{-0.4}$ \\ \hline
			Tracks $p$ measurable (MC) & 393 & 1 & 127 & 34 & 93 & 299 \\
			Tracks in selection domain (MC) & 2 & 0 & 9 & 0 & 13 & 19 \\
			Fraction (MC) [\%] & $0.5^{+0.7}_{-0.2}$ & $< 69.0$ & $7.1^{+3.0}_{-1.7}$
 & $< 6.4$ & $14.0 ^{+4.4}_{-2.9}$ & $4.8^{+1.3}_{-0.9}$ \\ \hline \hline  
		\end{tabular}
	\end{center}
\end{table}

\subsection{Goodness of agreement between the experimental and the simulated data}

The numbers of secondary particles in the domain where $\tau $ decay candidates are selected
are too few to evaluate the goodness of the agreement between the experimental and the simulated data.
So we try to widen the selection condition for comparison. 
We select 1-prong hadron interaction events with $\theta_{\rm kink} >$ 0.30, 0.15 and 0.06~rad
for 2~GeV/$c$, 4~GeV/$c$ and 10~GeV/$c$, respectively.
These angles correspond to $p_{\rm T}~>~0.6$~GeV/$c$.

There are some events in the domains defined by these conditions, 
as shown in Table~\ref{table:selected_data_vs_mc}.
We use these samples to check the agreement between the experimental and the simulated data.

In Table~\ref{table:selected_data_vs_mc}, numbers of simulated data (MC) were normalized 
to the total track length followed for each beam momentum in this experiment.
Then the statistical error $\delta _{\rm stat}$ and the relative difference $\delta _{\rm sys}$
are defined as follows,
\begin{eqnarray}
	\delta _{\rm stat} &=& \frac{1}{\sqrt{N_{\rm ev}}} \\
	\delta _{\rm sys} &=& \frac{N_{\rm ev} - N_{\rm MC}}{N_{\rm MC}} 
\end{eqnarray}
where $N_{\rm ev}$ and $N_{\rm MC}$ are numbers of the experimental data and the simulated data respectively.
It is found that all the relative differences $\delta _{\rm sys}$ are less than 30\% and 
can be understood within the statistical errors $\delta _{\rm stat}$
although statistics is not sufficient for 2~GeV/$c$.
We therefore conclude that the agreement between the experimental and the simulated data
has been confirmed at the 30\% level for beam momentum $P \geq 4$~GeV/$c$. 

\begin{table}[t]
	\caption[Numbers of selected events (experimental data and simulated data)]
{Number of selected events for each momentum beam.
We select 1-prong events with $\theta_{\rm kink} >$ 0.30, 0.15 and 0.06~rad
for 2~GeV/$c$, 4~GeV/$c$ and 10~GeV/$c$, respectively.
For the simulated data (MC), 
numbers of events normalized to the total track length followed in this experiment are shown. 
Relative differences $\delta _{\rm sys}$ between experimental data and simulated data
are less than 30$\%$ and are within statistical errors $\delta _{\rm stat}$ of experimental data
although statistics is not sufficient for 2~GeV/$c$.}
	\label{table:selected_data_vs_mc}	
	\begin{center}
		\begin{tabular}{cccccccccc} \hline \hline

			$P$~[GeV/$c$ ] & 2 & 2 (MC) & $\delta _{\rm sys}$ [$\%$] & 
4 & 4 (MC) & $\delta _{\rm sys}$ [$\%$]
& 10 & 10 (MC) & $\delta _{\rm sys}$ [$\%$]\\ \hline
			Events & 77 & 68.5 & 12 & 68 & 60.2 & 13 & 173 & 166.0 & 4 \\
			1-prong & 33 & 40.6 & -19 & 29 & 29.1 & 0 & 26 & 34.6 & -25 \\
			After $\theta_{\rm kink}$cut & 6 & 7.5 & -20 & 19 & 15.3 & 24 & 17 & 23.7 & -28 \\ \hline \hline
		\end{tabular}
	\end{center}
\end{table}

\section{Conclusions}
Topological and kinematical characteristics of hadron interactions have been studied 
by using an ECC brick exposed to 2, 4, 10~GeV/$c$ pion beams.
High speed automated microscope system was employed to analyze the nuclear emulsion films.
A total of 318 hadron interactions were found and reconstructed by 
following 60~m $\pi ^-$ tracks in the brick.
Secondary charged particle tracks from interaction vertices were also followed and reconstructed. 
Charged particle multiplicity of each event and emission angle of each secondary particle
were measured and their distributions were found to be in good agreement 
with a FLUKA Monte Carlo simulation.
Nuclear fragments were also searched for by 
newly developed automated microscope system with a wide view.
We measured the probability for the interaction vertices to be associated with nuclear fragments
and found it to be greater than 50\% for beam momentum $P > 4$~GeV/$c$.
The experimental data of the fragment association probability 
are well reproduced by the simulation with differences less than 10\%.
When possible, the momentum of the secondary particle was measured by using the coordinate method.
Fractions of 1-prong and 3-prong hadron interactions being in the domain
where $\tau$ $\to$ hadron decays are selected are measured and found to be consistent
with the simulation at the 30\% level for beam momentum $P \geq 4$~GeV/$c$.
We conclude that the FLUKA based simulation reproduces well the experimental distributions of 
2, 4, and 10~GeV/$c$ $\pi^-$ interactions in the ECC brick.
This result can be applied in neutrino oscillation experiments to evaluate the 
hadronic background contaminations on $\tau$ lepton decays.
Since the background contaminations originate from all charged hadron interactions,
mainly $\pi^-$, $\pi^+$ and proton interactions, positively charged hadron interactions
should also be investigated. This will be the subject of further studies.

\section*{Acknowledgments}
We wish to thank P. Vilain for his careful reading of the manuscript and for his valuable comments.
We wish to express our gratitude to the colleagues of the Fundamental Particle Physics Laboratory,
Nagoya University for their cooperation.
For the beam exposure, we gratefully acknowledge the support of the PS staff at CERN.
We warmly acknowledge the financial support from the Japan Society for the Promotion of Science (JSPS), 
the Promotion and Mutual Aid Corporation for Private Schools of Japan, and
Japan Student Services Organization.


\end{document}